\documentclass[a4paper,11pt]{article}
\usepackage{jcappub} 
\newcommand{\xmax}{\ensuremath{X_{max}\,}}
\usepackage[american]{babel}
\usepackage{amsmath}
\usepackage{amssymb}
\usepackage{graphicx} 
\usepackage{enumerate}
\usepackage{subcaption} 
\usepackage{color} 
\usepackage[normalem]{ulem}
\usepackage[T1]{fontenc}
\usepackage[utf8]{inputenc}

\title{Feasibility of event-by-event primary mass discrimination using radio observables and supervised machine learning}

\author[1]{Washington R. de Carvalho Jr.\note{Corresponding author.}}
\author{and Lech Wiktor Piotrowski}
 
\affiliation{Faculty of Physics, Warsaw University, ul. Pasteura 5, Warsaw, Poland}

\emailAdd{carvajr@gmail.com}

\abstract{In this work, we investigate the feasibility of event-by-event primary mass discrimination using radio observables only. Although the analysis does not require an explicit reconstruction of the shower maximum (\xmax), the discrimination power still arises from the sensitivity of the radio observables to the longitudinal development of the extensive air shower (EAS). Such radio-based approaches could be particularly relevant for radio-only experiments, such as GRAND. To assess this feasibility, we obtained conservative upper limits for the discrimination accuracy using a supervised machine-learning (ML) algorithm, namely a random forest (RF). The input features used were the peak electric fields and the spectral slopes, which have complementary discrimination power, along with the antenna distances to the shower axis. The RF was trained and tested using large event sets generated by the fast radio emission simulation and simplified detector response implemented in the RDSim framework. We obtained discrimination accuracies between 81\% and 96\% over the studied zenith range, even after normalizing each shower by its own electromagnetic energy. Since the analysis includes deliberately conservative choices, such as a large 10\% uncertainty on the reconstructed EM energy, these quoted values should be interpreted as conservative upper limits suitable for a feasibility assessment. Our results demonstrate that event-by-event primary mass discrimination using radio observables is, in principle, feasible.}

\begin{document}
\maketitle
\flushbottom

\section{Introduction}
\label{sec:intro}

One of the main objectives of cosmic ray (CR) research is to identify the sources of ultra-high-energy cosmic rays (UHECRs). However, due to the potentially large deflections of charged primary particles in the Galactic magnetic field (GMF) during propagation to Earth, inferring the source direction from the arrival direction of a shower is extremely challenging in most cases. However, for the most energetic events, deflections can be significantly smaller. Therefore, cosmic-ray backtracking techniques \cite{Farrar2013, Farrar2019}, which utilize detailed GMF models \cite{Jansson2012, UngerFarrar2024}, can be used to obtain a more constrained estimate of the source direction of such events. While extragalactic magnetic fields (EGMF) also contribute to the total deflection \cite{AlvesBatista2017,Hackstein2018}, stochastic energy losses in intergalactic space prevent true backtracking, and forward-tracking techniques must be used instead. Since the magnitude of deflections strongly depends on the particle’s rigidity, an estimate of the event's primary mass is required. This makes the development of event-by-event mass reconstruction methods crucial for establishing the sources of the highest-energy CRs.

Determining the mass composition of CR events is one of the most challenging tasks for UHECR experiments. Traditional analyses typically estimate the average mass or mass fractions for a given CR flux, rather than performing event-by-event mass reconstructions. Also, as is customary in the field, \xmax is commonly used as a surrogate for mass composition, since it is highly correlated with the type of particle that initiated the extensive air shower (EAS). In this context, we present a machine learning (ML) approach based on classification random forests (RF) for event-by-event primary mass discrimination of cosmic ray events. Each event is assigned to one of two classes: heavy, corresponding to iron-like showers, or light, corresponding to proton-like showers. Unlike conventional methods, this approach bypasses \xmax reconstructions and instead directly infers the primary composition of each event. Conceptually, this ML method resembles a previous one we have developed \cite{composition-ARENA2018,compositionpaper}, as both perform event-by-event analyses without reconstructing \xmax. Rather than ML, this earlier method relied on a $\chi^2$ comparison of each event with multiple simulations of different compositions, akin to LOFAR-like \xmax reconstruction methods, but it stops short of reconstructing \xmax and infers the primary composition directly. One should note that it is the position of \xmax in the atmosphere that largely determines the characteristics of the electric field at ground level. So, although neither method reconstructs \xmax, it is the position of the shower maximum, not the primary composition itself, that enables our event-by-event discrimination.

An interesting aspect of supervised ML algorithms, such as RFs, is that analyzing feature importances can reveal which shower characteristics drive the discrimination. This approach motivated a separate study~\cite{paperRLDF,RLDF-ICRC2025}, in which we explore the dependence of electric field amplitudes on \xmax and the underlying physics. In addition to the peak electric field amplitude, the spectral slope of the radio signal at each antenna has also been shown to correlate with \xmax and thus with the primary mass of the cosmic ray~\cite{Jansen2016,Canfora2021}. In our ML discrimination method we use both the amplitude and the spectral slope, as they offer complementary sensitivity to composition.

ML methods require very large data sets for training and testing, which makes the use of a dedicated full Monte Carlo simulation of the radio emission for each event, such as ZHAireS \cite{zhaires-air}, computationally prohibitive. For this reason, the event data sets used in this work were generated with RDSim \cite{RDSim-ARENA2022,RDSim-ECRS,RDSim-ICRC2023}, a fast yet comprehensive Monte-Carlo tool that models the radio emission and its detection at a given antenna array. RDSim allows the morphing of a single full ZHAireS input simulation to generate multiple realistic events with different core positions, arrival directions, and energies. While RDSim can rotate a shower to simulate different arrival directions and estimate the radio signal at any antenna position, it does not alter the underlying shower development. Therefore, we still rely on many independent ZHAireS simulations to account for shower-to-shower fluctuations.

In this work, we probe the feasibility of event-by-event primary mass discrimination using supervised machine learning, focusing on the potential of radio-based observables to separate light and heavy primaries. This paper is organized as follows. In Section \ref{sec:Simulations}, we describe the ZHAireS simulations used as input for this work. Section \ref{sec:RDSim} presents the RDSim framework, including the superposition emission model it uses to estimate the electric field amplitude and spectral slope at any position. In section \ref{sec:events} we describe the events generated for this study, along with the parameters used. Section \ref{sec:RF} describes the ML mass discrimination method, including the RF features and parameters. Section \ref{sec:Results} presents the discrimination results and an analysis of the feature importances, which we use to understand how the RF leverages different features across zenith ranges. Finally, Section \ref{sec:Discussion} provides a discussion of the results and their implications, followed by the conclusions in Section \ref{sec:Conclusions}. In the Appendix, we include more in-depth information on RDSim, including comparisons of its results with full-fledged simulations of both the radio emission and detector response.

\section{Input ZHAireS simulations}
\label{sec:Simulations}

The radio emission from EAS is produced by the motion of charged particles in the atmosphere, which can be interpreted as currents. Given an observer, the amplitude of the induced electric field is roughly proportional to the projection of these currents in the direction perpendicular to the observer. This perpendicular current interpretation forms the basis of the ZHS formalism~\cite{ZHS92,TimeDomainZHS}, which is used by ZHAireS~\cite{zhaires-air} to compute the net electric field by summing the contributions from all individual particle tracks in the shower. Macroscopically, the radio emission from EAS can be seen as arising from two main emission mechanisms, Askaryan and geomagnetic. The geomagnetic emission arises from the opposite deflections of electrons and positrons in the geomagnetic field, leading to an emission polarized along the $\vec{V} \times \vec{B}$ direction, where $\vec{V}$ is the direction of the shower axis and $\vec{B}$ is the geomagnetic field (see left panel of Fig.~\ref{fig:toymodelscheme}). The Askaryan, or charge-excess, emission arises from an excess of electrons in the shower front and is radially polarized with respect to the shower axis. The superposition of the Askaryan and geomagnetic mechanisms, each with its distinct polarization, creates the radio footprint at ground level. Since the Askaryan polarization depends on the observer position, while the geomagnetic does not, this footprint is asymmetric.

ZHAireS~\cite{zhaires-air} is a Monte Carlo simulation code that combines the AIRES~\cite{aires} shower simulation framework with the ZHS~\cite{ZHS92,TimeDomainZHS} algorithm. It applies the ZHS formalism to the charged particle tracks simulated by AIRES to compute the radio emission from the EAS. The radio signal is calculated from first principles and does not presuppose any specific emission mechanism. Nevertheless, the net electric field obtained from ZHAireS is compatible with a superposition of the Askaryan and geomagnetic contributions. In this work, the full ZHAireS simulations are not used directly. Instead, they provide the input for the superposition radio emission model in RDSim, which in turn generates the events used in the analysis. This model is described in detail in Section~\ref{sec:toymodel}.

We performed ZHAireS simulations at the site of the Giant Radio Array for Neutrino Detection (GRAND) in China, using a ground altitude of 1264~m~a.s.l. and a geomagnetic field of intensity $|\vec{B}|=56.4~\mu$T with an inclination of $61.6^\circ$. All showers were set to come from the North, with zenith angles from $42^\circ$ to $82^\circ$, in steps of $4^\circ$. Approximately 90 antennas were placed on a single line, East of the shower core. For each zenith, we simulated a total of 50 p and 50 Fe showers and varied the antenna spacing to obtain the electric field at up to $\sim4.5$ times the expected Cherenkov ring distance. The primary energy was set to $E_0=1.25$~EeV and the SIBYLL hadronic model~\cite{SIBYLL} was used. Each shower was simulated twice with the same seed: once with $\vec{B}$ on and once off (see Section~\ref{sec:toymodel}). We then used these simulations to create a total of 100 instances of the superposition radio emission model per zenith in RDSim, after applying a bandpass filter between 30 and 80~MHz. These simulations are exactly the same as those described in \cite{RLDF-submitted} for the GRAND site.

For each shower, we have also obtained an estimate of its EM energy by analyzing the particle stack of the simulation. This includes the deposited energy of all EM particles as well as the energy of discarded or low-energy particles. The resulting EM/missing energies are consistent with values reported in the literature~\cite{AugerInvisibleEnergy,ICRCBeijingMissingEnergy,MatiasMissingEnergy,KASCADEMissingEnergy}. It is well known that the radio emission of air showers depends on the EM energy of the shower, not on the primary energy. For a set of showers with the same primary energy $E_0$ but different compositions, the field amplitude would not only depend on the shower development (\xmax), but also on the EM/hadronic energy ratio of the shower, which tends to decrease with primary mass. This means that if $E_0$ were known, this amplitude dependence on the EM/hadronic ratio could in principle be used to further improve primary mass discrimination. However, energy reconstructions that use the radio technique actually measure the EM energy of the shower, not $E_0$. A set of events in an energy bin reconstructed with such methods would thus have the same average EM energy, with a dispersion that reflects the uncertainty of the reconstruction. For this reason, we normalize each ZHAireS simulation by its own EM shower energy, before using it as input for RDSim. All ZHAireS simulations now have the same EM energy instead of the same $E_0$, partially mimicking a radio reconstructed energy bin. To further mimic real events, a dispersion in EM energy will be included later, when generating each event with RDSim.

\section{RDSim}
\label{sec:RDSim}

RDSim~\cite{RDSim-ARENA2022, RDSim-ECRS, RDSim-ICRC2023} is a fast, accurate, and comprehensive framework for the simulation of the radio emission from EAS and its detection by arbitrary antenna arrays. The radio emission component of RDSim is based on a superposition model that uses full ZHAireS simulations as input and disentangles the geomagnetic and Askaryan contributions, building on the approach introduced in~\cite{toymodel}. Over the years, this superposition model has been systematically expanded for accuracy and additional capabilities. By assuming an elliptical symmetry at ground, it is able to estimate the peak electric field and its polarization, as well as the spectral slope of the signal, at any observer position on the ground. At the detector level, the framework takes into account the main characteristics of the detector, such as the antenna array layout, trigger setups, thresholds, and antenna patterns.

RDSim is capable of reusing a single full ZHAireS simulation to generate multiple realistic events with different arrival directions, energies, and core positions at a very low computational cost. That, coupled with the efficient modeling of the detector response, makes RDSim extremely fast, allowing it to generate huge datasets covering a wide range of geometries and energies in a very short time. As a result, RDSim is particularly well suited to produce the large statistics needed to train and test ML algorithms, as was done in this work.

\subsection{Superposition model}
\label{sec:toymodel}

RDSim uses the two full ZHAireS simulations available for each event (with $\vec{B}$ enabled and disabled) to disentangle the Askaryan and geomagnetic contributions at every antenna along the reference line (i.e., the line where antennas were fully simulated). We then use Hilbert envelopes to obtain the peak amplitude as a function of distance to the core, for each mechanism separately. This same reference line will also be used to store the spectral slope as a function of distance to the core (see Section~\ref{sec:slope}).

\begin{figure}[!htb]
  
  \begin{center}
    
    
    \begin{minipage}{.49\textwidth}
      \hspace{1cm}\includegraphics[width=0.75\linewidth]{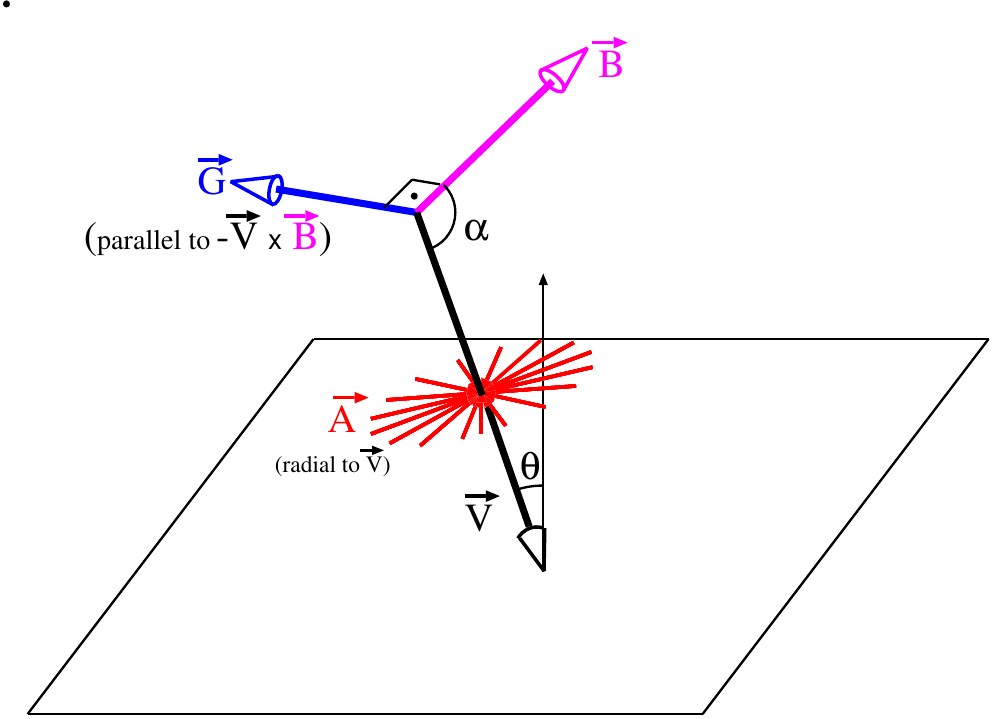}
    \end{minipage}
    \begin{minipage}{.49\textwidth}
      \vspace{0.6cm}
      \includegraphics[width=0.86\linewidth]{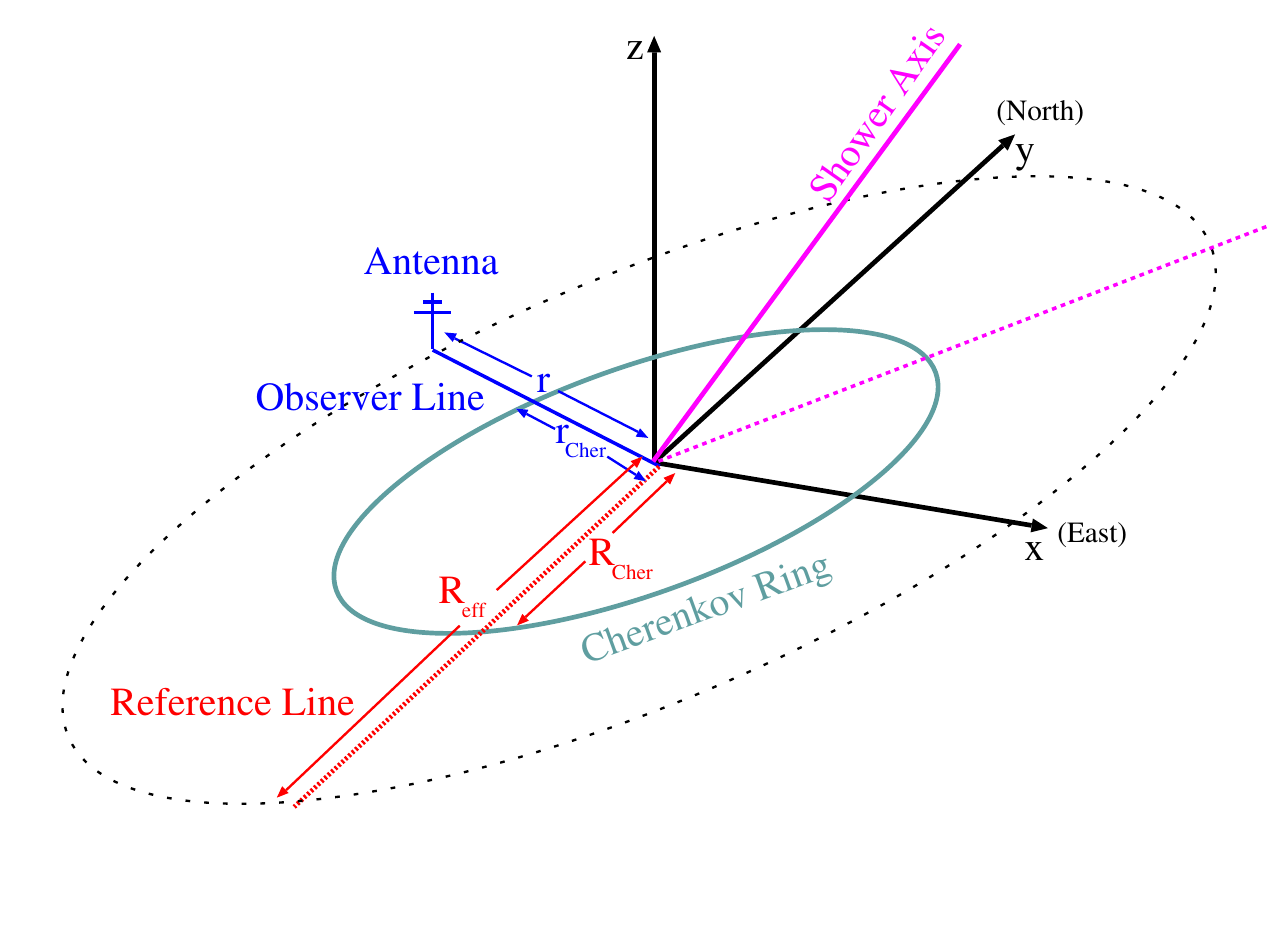}
    \end{minipage}
  \end{center}

  
  \caption{Left: Theoretical polarization directions of the Askaryan (red) and geomagnetic (blue) emission mechanisms. $\vec{V}$ (black) denotes the shower axis and $\vec{B}$ (magenta) the geomagnetic field. Right: Schematic of the elliptical symmetry at ground level used by the superposition emission model. The reference line, where full simulations are performed, is shown in red, while the antenna position for which the electric field is estimated is shown in blue. Both panels are adapted from \cite{toymodel}.}
  \label{fig:toymodelscheme}
\end{figure}

Using the position of \xmax, the shower geometry, and a model of the atmosphere, we then calculate the Cherenkov angle at \xmax, thus defining the Cherenkov cone. The intersection of this cone with the ground defines the Cherenkov ellipse, where the radio signal is expected to be maximal. Any observer located on the Cherenkov ellipse views \xmax at the same angle with respect to the shower axis. In fact, this angular symmetry also applies to any similar ellipse, i.e., any ellipse with the same ratio between the major and minor axes as the Cherenkov ellipse. The superposition model exploits this angular (elliptical) symmetry by assuming that the Askaryan and geomagnetic amplitudes are constant along a given similar ellipse. For any arbitrary observer position on the ground (blue antenna on the right panel of Fig.~\ref{fig:toymodelscheme}), the corresponding similar ellipse is first identified (dashed ellipse on the figure) using the direction and distance $r$ from the core to the antenna. We then find the intersection of this ellipse with the reference line (red line), defining an effective distance $R_{\mbox{eff}}$ to the shower core that corresponds to the same ellipse as the observer position. The Askaryan and geomagnetic amplitudes are sampled at this point and the net peak electric field and polarization are obtained by combining these sampled amplitudes with the theoretical polarization direction for each emission mechanism at the observer position (see left panel of Fig.~\ref{fig:toymodelscheme}). More details about these procedures can be seen in~\cite{toymodel}.

To enhance both accuracy and flexibility, the original superposition model has been extended in several ways. One such extension is the inclusion of an early-late correction. The amplitudes sampled along the reference line are now scaled by the factor $D_{\mbox{eff}}/D_{\mbox{obs}}$, where $D_{\mbox{eff}}$ is the distance between \xmax and the reference line sampling point, and $D_{\mbox{obs}}$ the distance from \xmax to the desired antenna position. This correction essentially implements the expected $1/R$ distance scaling of the electric field, which becomes particularly important for showers at large zenith angles. Another extension allows the emission model to be rotated freely in azimuth. In this procedure, the entire model geometry is rotated to the new arrival direction, including the shower axis, the Cherenkov ellipse and the reference line. The geomagnetic amplitudes along the (now rotated) reference line are then rescaled to reflect the change in the angle $\alpha$ between the shower axis and the geomagnetic field by assuming the theoretical $\sin(\alpha)$ scaling of the geomagnetic component of the electric field. This rotation procedure allows the model to calculate the radio emission of showers arriving from different azimuth angles and makes it possible to use a single ZHAireS simulation to generate events for multiple arrival directions, highlighting the versatility and efficiency of the framework.

\begin{figure}[!htb]
  \begin{center}
    \includegraphics[width=0.65\linewidth]{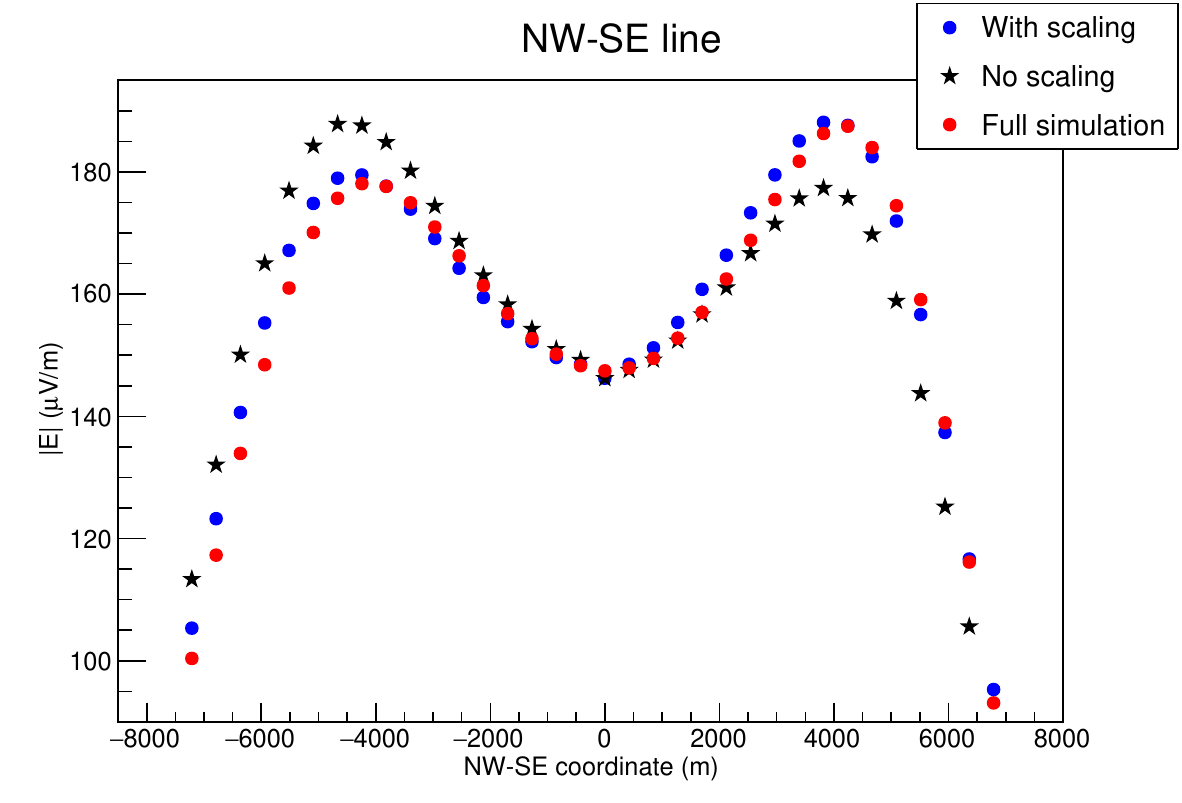}
  \end{center}


  \caption{Comparison between full ZHAireS simulations and the results of the superposition model for the peak electric field amplitudes. Shown are the amplitudes for an $80^\circ$ shower along the major axis of the elliptical radio footprint. The curves labeled ``With Scaling'' (``No Scaling'') correspond to results with (without) the early-late correction. Including this correction brings the superposition model into very good agreement with the full simulation, with a maximum difference of about 6\% in this example.}
  \label{fig:compamplitude}
\end{figure}

The superposition model implemented in RDSim has been extensively validated against full ZHAireS simulations across a wide range of shower geometries. An example comparison can be seen on Fig.~\ref{fig:compamplitude}, where we compare the peak electric field amplitudes obtained from full ZHAireS simulations with results from the superposition model  for an 80$^\circ$ shower. The amplitudes are shown along the major axis of the elliptical radio footprint, where early-late effects reach their maximum. The labels ``with scaling'' (``no scaling'') refer to results with (without) the early-late correction. One can see that the early-late correction brings the model into very good agreement with the full simulation, with a maximum difference of $\sim6$\%, even in this high zenith example. We have also analyzed the errors related to the rotation of the emission model. As can be seen in Appendix~\ref{sec:A-RotationExample}, these errors are even smaller, around 2\% in our example comparison of a rotated model against a non-rotated one constructed from a dedicated simulation. We have established the accuracy of the peak amplitudes obtained by the model. In the next section we will focus on the spectral slope of the radio emission.

\subsection{Spectral slope}
\label{sec:slope}

As in the case of the peak amplitudes, the emission model can estimate the spectral slope of the radio emission at any position on the ground, using the same ZHAireS simulations as input. Although the superposition of the Askaryan and geomagnetic components is not used to calculate the spectral slope, the emission model exploits the same angular (elliptical) symmetry described in Section~\ref{sec:toymodel}. It assumes that observers who see \xmax at the same angle relative to the shower axis measure the same spectral slope. This approach allows the model to efficiently reproduce the spatial distribution of the spectral slope across the ground while maintaining consistency with the reference simulations.

The spectral slope at each reference antenna is obtained by fitting the logarithm of the frequency spectrum of the fully simulated electric field with a quadratic function (see Appendix~\ref{sec:A-SSFit} for details). These values are stored as a function of distance to the shower core along the reference line (see Fig.~\ref{fig:toymodelscheme}), in the same way as the previously described Askaryan and geomagnetic amplitudes. For an arbitrary observer on the ground, the spectral slope is sampled at the corresponding point on the reference line, determined using the same effective distance $R_{\mbox{eff}}$ procedure as in Section~\ref{sec:toymodel}. This sampled value is then assigned to the observer.

\begin{figure}[!htb]
  \begin{center}
    \includegraphics[width=0.48\linewidth]{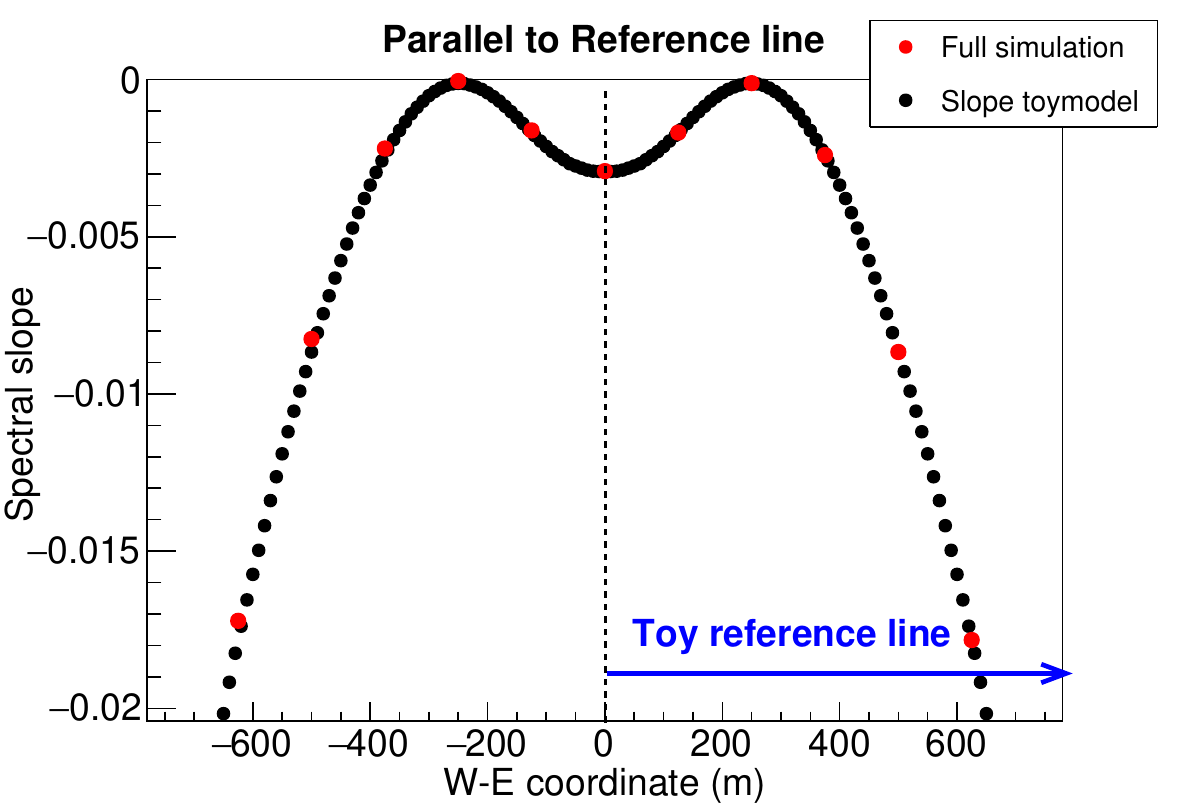}
    \includegraphics[width=0.48\linewidth]{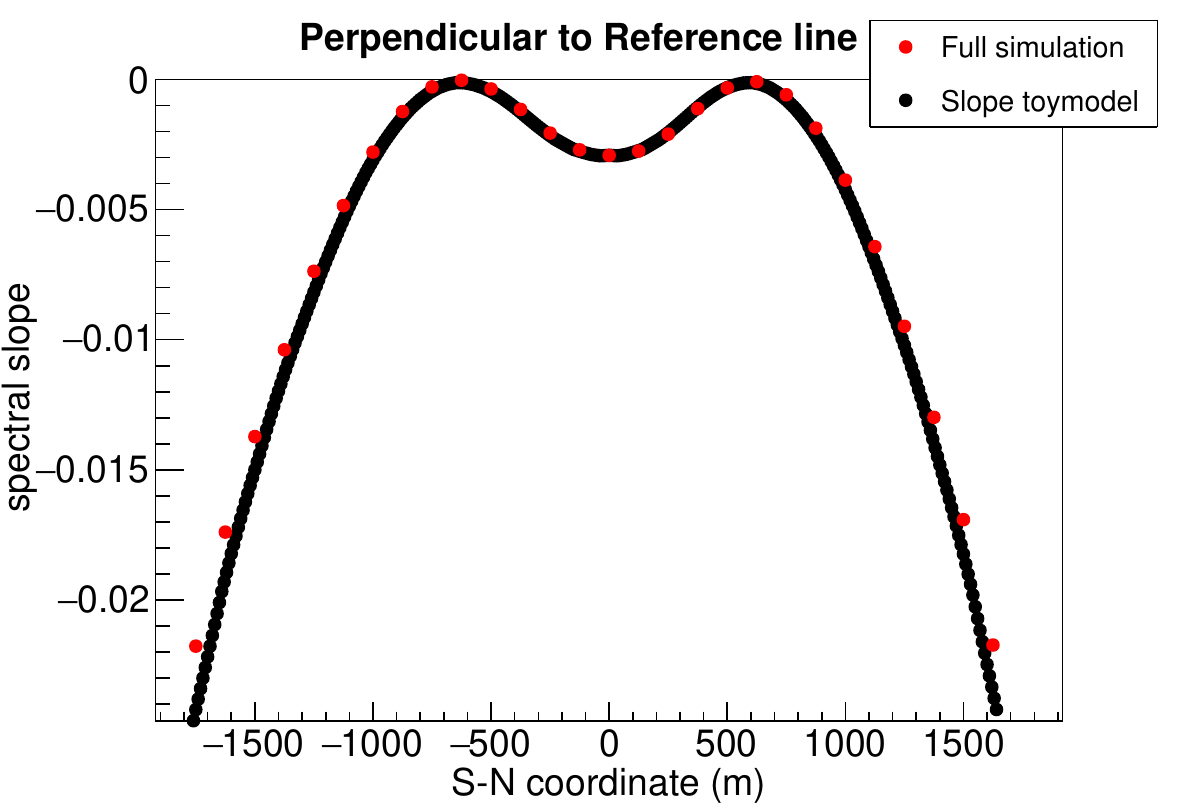}
  \end{center}


  \caption{Example comparison between full ZHAireS simulations and the results of the emission model for the spectral slope of the radio emission. Shown are the spectral slopes for a $66^\circ$ shower along the EW (left) and NS (right) directions. In both cases, the emission model shows very good agreement with the full simulation. The blue arrow indicates the position of the reference line.}
  \label{fig:compslope}
\end{figure}

Figure~\ref{fig:compslope} shows a representative comparison between spectral slopes obtained from full ZHAireS simulations and those predicted by the emission model for a $66^\circ$ shower, along the EW (left panel) and NS (right panel) directions. As can be seen in the figure, the emission model reproduces both the overall spatial structure and the absolute values of the spectral slope very well. The very good agreement observed in this comparison demonstrates that the model captures the dominant features governing the frequency dependence of the radio signal and that the angular symmetry used is valid not only for the amplitudes, but also for the spectral slope. We find good agreement between the model predictions and full simulations over the full zenith range used in this work. This indicates that the emission model can provide a reliable description of the spectral slope for the present ML application\footnote{Since the extension of the emission model to estimate the spectral slope is relatively recent, it has not yet been validated as extensively as the amplitude calculation. However, we have performed multiple comparisons with full simulations in the zenith range relevant for this work and found the level of agreement achieved to be more than adequate for the ML application presented here.}.

\subsection{Detection simulation}
\label{sec:detection}

RDSim was built from the ground up with speed in mind. Therefore it uses a lightweight yet accurate and fully configurable detection module that captures the essential characteristics of a radio array while remaining computationally efficient. It can be used with any arbitrary antenna layout, provided all antennas are on a horizontal plane at ground altitude. At the antenna level, triggers are modeled by applying a user defined electric field threshold, either to the net field or separately to each component. RDSim also provides a vertical polarization toggle that allows the triggering to be evaluated using only the horizontal polarization, matching the chosen antenna characteristics. Array level triggers require a user defined minimum number of triggered antennas for the whole event to be considered triggered.

RDSim can also take arbitrary antenna beam patterns into account in the detection simulation. When beam patterns are enabled, the peak electric field predicted by the superposition emission model is multiplied by the corresponding gain for the arrival direction (relative to \xmax) before the trigger simulation. Note that since the emission model only provides peak amplitudes, RDSim does not propagate any full electric field time traces. For that reason, the spectral slope obtained from the emission model is not passed through any detector response simulation. In a previous work~\cite{RDSim-ICRC2023}, we directly compared events generated with the full RDSim chain, which combines the superposition emission model with the detection module, to events produced using full radio emission and detector response simulations. To generate these reference events, we passed the electric field time traces obtained from full ZHAireS simulations through the complete Auger Offline detector response pipeline\footnote{Offline is the official simulation and reconstruction framework of the Pierre Auger Observatory.}. Three such comparisons from \cite{RDSim-ICRC2023} are reproduced in Appendix~\ref{sec:A-Offline} for completeness. All models in RDSim are intentionally simple, leading to computational costs that are orders of magnitude lower than full simulations. Despite this, the agreement with full end-to-end simulations is exceptionally good, far better than what one would typically expect from such simple models. This shows that RDSim provides a very reliable description of the radio detector response for the purposes of this work.

\section{Event generation for ML analysis}
\label{sec:events}

In this section we describe the parameters used by RDSim to generate the events for the ML analysis. As discussed earlier, RDSim takes the EM-normalized ZHAireS simulations as input to produce these events. As the detector, we used a generic hexagonal antenna array with 500 m spacing.\footnote{For $\theta = 82^{\circ}$ the antenna spacing was increased to 1000 m to reduce the number of triggered antennas and speed up training. Note that the radio footprint remains very well sampled at this zenith, even with this increased spacing.} We used a large enough array to ensure that all events would be completely contained. Also, given the generic nature of the array, no specific antenna beam pattern was applied. The trigger model used all components of the electric field, with an antenna-level threshold of 101~$\mu$V/m per component. Each event required a minimum of five triggered antennas to be accepted.

For each zenith angle we generated $\sim$10k events, which were then divided into two statistically independent samples of $\sim$5k events each: one for training the random forest and the other for testing its accuracy. The shower azimuth was sampled equally over the full range, and the core position was sampled uniformly in area, in a region that ensured full containment of the footprint. A Gaussian energy smearing of 10\% was then applied to each event, which is twice the $\sim$5\% uncertainty quoted for modern radio-based EM energy reconstructions (e.g.\ \cite{FelixTimInclinedRec}). This very conservative choice ensures that the energy fluctuations are not underestimated, since they directly affect the electric field amplitude feature used by the RF. By adopting an uncertainty much larger than expected, we make sure that we are not overestimating the discrimination power. If anything, we are likely underestimating it. This energy smearing, combined with the previous EM energy normalization performed on the ZHAireS input simulations, makes our events closely resemble those in a single energy bin reconstructed with radio-based methods.

As a final note, remember that all intrinsic shower-to-shower fluctuations in our events originate from the set of 100 ZHAireS input simulations per zenith angle, as RDSim only modifies the shower geometry and does not alter the shower development itself. Consequently, even when generating 10k events, the resulting sample contains only 100 distinct values of \xmax. We consider this level of sampling sufficient to capture the main characteristics of shower-to-shower variability for the purpose of this study. An illustrative example of a simulated event, showing the resulting footprint and slope pattern, is provided in Appendix~\ref{sec:A-EventExample}.

\section{ML mass discrimination method}
\label{sec:RF}

In this section we describe the supervised ML method used for the event-by-event binary discrimination between light (p-like) and heavy (Fe-like) cosmic-ray primaries.
The classification is performed by a random forest classifier implemented in a custom general-purpose code, using exclusively radio observables measured at the triggered antennas. The purpose of this analysis is to assess, in a controlled Monte Carlo environment, the feasibility of a radio event-by-event mass discrimination method that bypasses any explicit reconstruction of \xmax. Nevertheless, as previously stated, the discrimination power still depends on the shower development.

The RF uses vectors of observables and contains no information on air-shower physics or any explicit composition dependence. All physical information is provided exclusively through the input feature vector that contains, for each triggered antenna, the distance to the shower axis ($D_{Ai}$), the peak electric field amplitude ($|E_i|$), and the spectral slope of the radio signal ($SS_i$). The triggered antennas of each event are sorted by increasing distance to the shower axis, and their observables are then appended sequentially to build the feature vector, $(D_{A1}, |E_1|, SS_1;\; D_{A2}, |E_2|, SS_2;\; \ldots;\; D_{AN}, |E_N|, SS_N)$, where index 1 denotes the antenna closest to the axis, index 2 the second closest, and so on. The total number of features is given by $N_f = 3N$, with $N$ defined as the number of triggered antennas in the highest multiplicity event. For events with fewer triggered antennas, the missing entries are set to zero (zero padding), in order to keep the input dimensionality fixed. In most cases, due to the presence of lower multiplicity events, the last features are dominated by zeros. For this reason, we included an option in the RF to disregard the last $N_R$ entries in order to reduce the numerical overhead and improve stability.

For each zenith angle, the RDSim generated events (see Section~\ref{sec:events}) are divided into statistically independent training and test sets. Separate random forests, composed of a fixed number of decision trees, are independently trained for each zenith angle. Each individual tree uses a bootstrap sample of the training data, where events are drawn from the full training set with replacement. As a result, the same event can appear more than once in the training sample of a given tree, while other events may not be used at all for that particular tree. During the construction of each tree, only a randomly selected subset of the available features is considered when determining the optimal split at a given node. The size of this subset is treated as a tunable hyperparameter. In this work, the empirically obtained values were found to be well above the commonly used $\sqrt{N_f}$ prescription. A maximum tree depth is imposed as a safety bound, but in practice it was never reached in this analysis. Instead, the effective stopping criterion is provided by the requirement of a minimum number of samples per leaf. These choices were made to improve stability and avoid over-fitting.

In this work, we use a general-purpose custom RF implementation that we previously developed. It is a pure numerical classifier, without any built-in physical knowledge, such as air-shower physics, detector geometry, or mass composition. The code only uses as input a vector of features, which in this particular application is constructed from radio observables. Our implementation follows standard, well-established RF algorithms and can use both the Gini impurity and the entropy criteria for node splitting. In this analysis we only use the Gini metric, since it yielded slightly better results. It also includes several general-purpose tools that are particularly useful for this analysis, such as feature-importance calculation using the shuffle method, flexible handling of zero-padded features, multi-thread support, and binary I/O of trained forests along with importances and other diagnostic outputs. These tools remain fully generic and do not encode any air-shower-specific knowledge. The RF code has been validated on several independent benchmark classification problems with known solutions. In all cases, the expected classifications were correctly recovered. This validation was performed prior to its first application to air-shower data. All this ensures that the results reported in this work reflect the physical information content of the input observables rather than any artifacts of the RF implementation.

Our code uses the shuffle method to calculate feature importances. In this method, the values of a given feature are randomly permuted across the events, and the corresponding degradation of the discrimination accuracy is measured. This degradation is then used as a relative measure of the importance of that feature for the RF decision. Since the input features can be strongly correlated, permutation-based importances may be redistributed among correlated features, particularly in regimes where multiple observables contain comparable physical information. In such cases, the resulting importances must be interpreted with care, primarily at the level of broader feature regions rather than as strictly local single-feature quantities. In Section \ref{sec:Results} we examine feature importance patterns at different example zenith angles. This can help us understand which observables the RF effectively relies on for the p–Fe discrimination.

No systematic hyperparameter optimization was performed in this work. During early development we carried out empirical explorations to identify stable working ranges for the most relevant RF parameters. The $|E|$-only configuration received the most extensive early testing, whereas the SS-only case was explored more lightly. For the combined $|E|$+SS configuration, which is the primary result of this study, no dedicated tuning was performed and all RF parameters were fixed to values previously validated as robust. All forests used in this work contain 200 trees, a maximum depth of 100 (never reached), and a minimum of 10 samples per leaf; these values were kept fixed throughout. The feature subset size varied with zenith angle, and values around $\sim 0.8\,N_f$, well above the conventional $\sqrt{N_f}$ prescription, consistently provided the most reliable performance. This behavior is consistent with the strong correlations expected among the input features, which all sample the same radio footprint. The bootstrap size was also varied during the exploratory tests, with larger fractions generally yielding better performance but at increased computational cost. However, for the final $|E|$+SS configuration, a small fixed bootstrap fraction of $\sim 0.2\,N_E$ was adopted to reduce computation time while maintaining stable results, with $N_E\simeq5000$ events per training set (see Sec.~\ref{sec:events}). In all cases, small values of the truncation parameter $N_R$ yielded the best performance, since low-multiplicity events lead to zero-padding in the last features. Because we performed no systematic optimization of the RF hyperparameters and intentionally used a conservative energy uncertainty that is twice the expected experimental value (see Sec.~\ref{sec:events}), the reported discrimination accuracies should be regarded as conservative estimates. Further improvements may be achievable with a dedicated optimization campaign, but this lies beyond the scope of the present feasibility study, particularly because we do not include additional detector-level uncertainties beyond the EM energy uncertainty, which directly impacts the radio-signal amplitude (see Sec.~\ref{sec:Discussion}).

\section{Results}
\label{sec:Results}

In this section we present the event-by-event mass discrimination accuracies obtained with our RF classifier using radio observables. These results quantify the performance of the method and are the main focus of this work. The classification accuracy is defined here as the fraction of correctly classified events in the statistically independent test sample. To assess the performance of each observable on its own, we first examine the configurations that use either the peak electric field or the spectral slope. The amplitude only configuration was the first to be investigated and motivated a previous study that examined the connection between \xmax and peak amplitudes \cite{RLDF-submitted}. We then consider the combined configuration, which uses both observables simultaneously and provides the main metric for assessing the feasibility of event-by-event mass discrimination.

\begin{figure}[!htb]

  
  \begin{center}
    \includegraphics[width=0.64\linewidth]{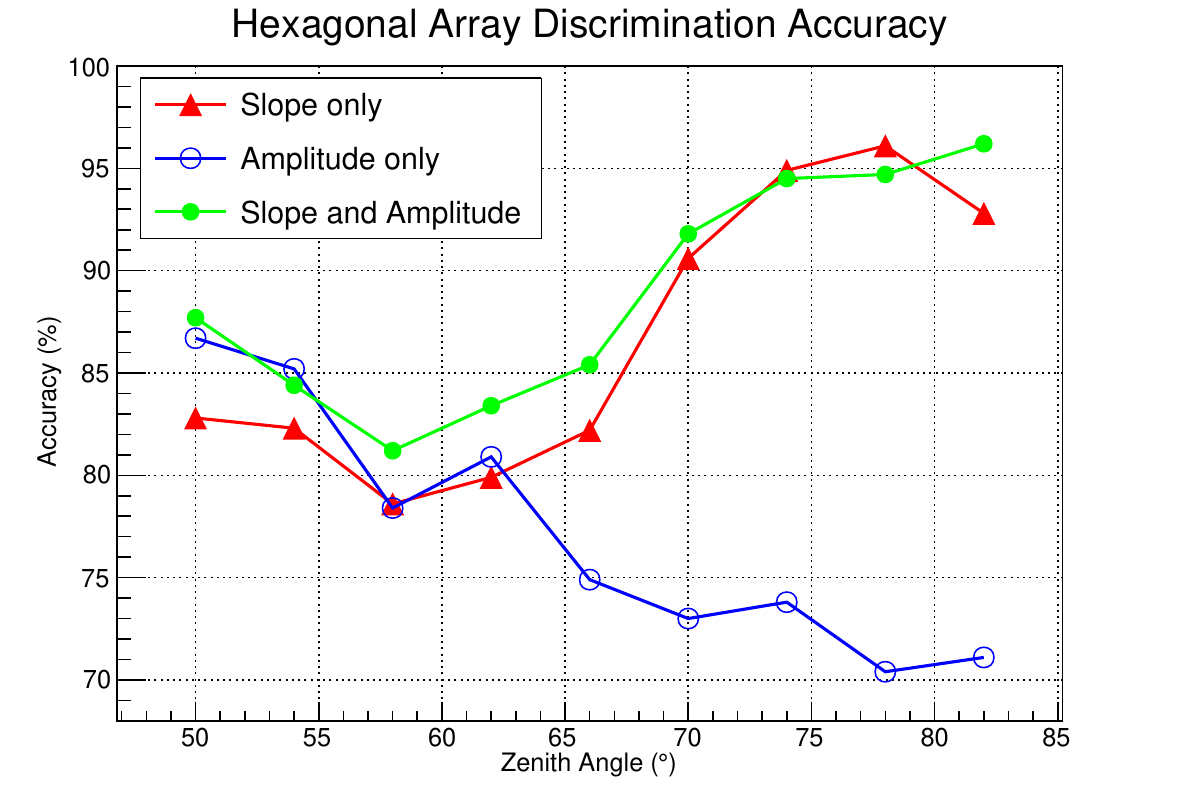}
  \end{center}
  
  
  \caption{RF discrimination accuracy vs zenith angle. See text for more details.}
  \label{fig:Results}
\end{figure}

Figure~\ref{fig:Results} shows the discrimination accuracies as a function of zenith angle obtained using the three studied configurations\footnote{The $|E|$-only curve was obtained using earlier RF trainings performed with a 50--200\,MHz bandwidth. In this work, all configurations involving the spectral slope use a 30--80\,MHz bandwidth. The $|E|$-only results are included here for completeness.}, for events normalized by their electromagnetic shower energy and including a 10\% Gaussian EM-energy smearing: $|E|$-only (blue empty circles), $SS$-only (red filled triangles) and $|E|+SS$ (green filled circles). The $|E|$-only accuracy is highest at low zenith and tends to decrease with increasing zenith. The $SS$-only accuracy decreases slightly from $50^\circ$ to a minimum near $58^\circ$ and then increases rapidly toward its maximum at the highest zenith angles. These almost opposite behaviors suggest that the two observables carry complementary information: while the amplitude provides its highest discrimination power at low zenith, the spectral slope achieves its maximum at the highest zenith angles. When both $|E|$ and $SS$ are used simultaneously (green line), the RF generally reaches higher accuracies than when either is used alone, as expected. While at high zenith the combined performance is similar to the $SS$-only case, suggesting that $SS$ dominates in this region, at low to mid zenith the RF appears to benefit from the information carried by both observables.

\begin{figure}[!htb]

  
  \begin{center}
    \includegraphics[width=0.5\linewidth]{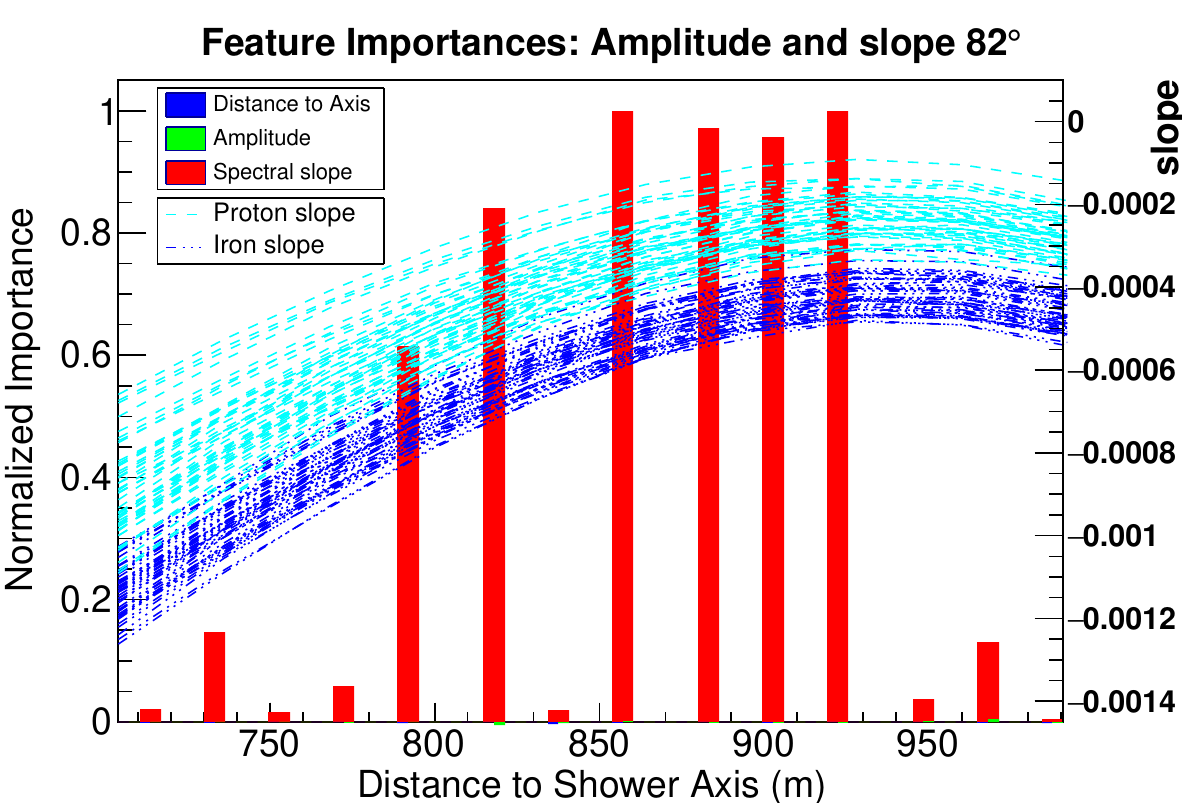}
    \includegraphics[width=0.5\linewidth]{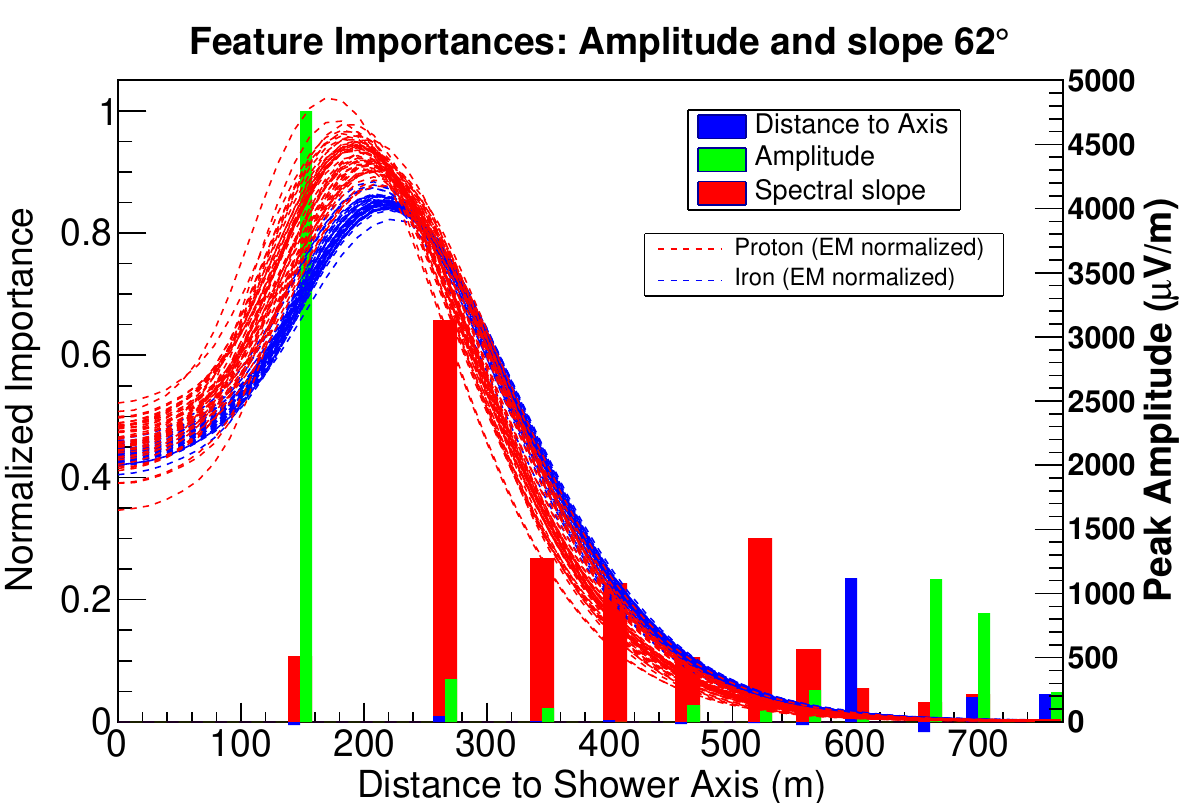}\includegraphics[width=0.5\linewidth]{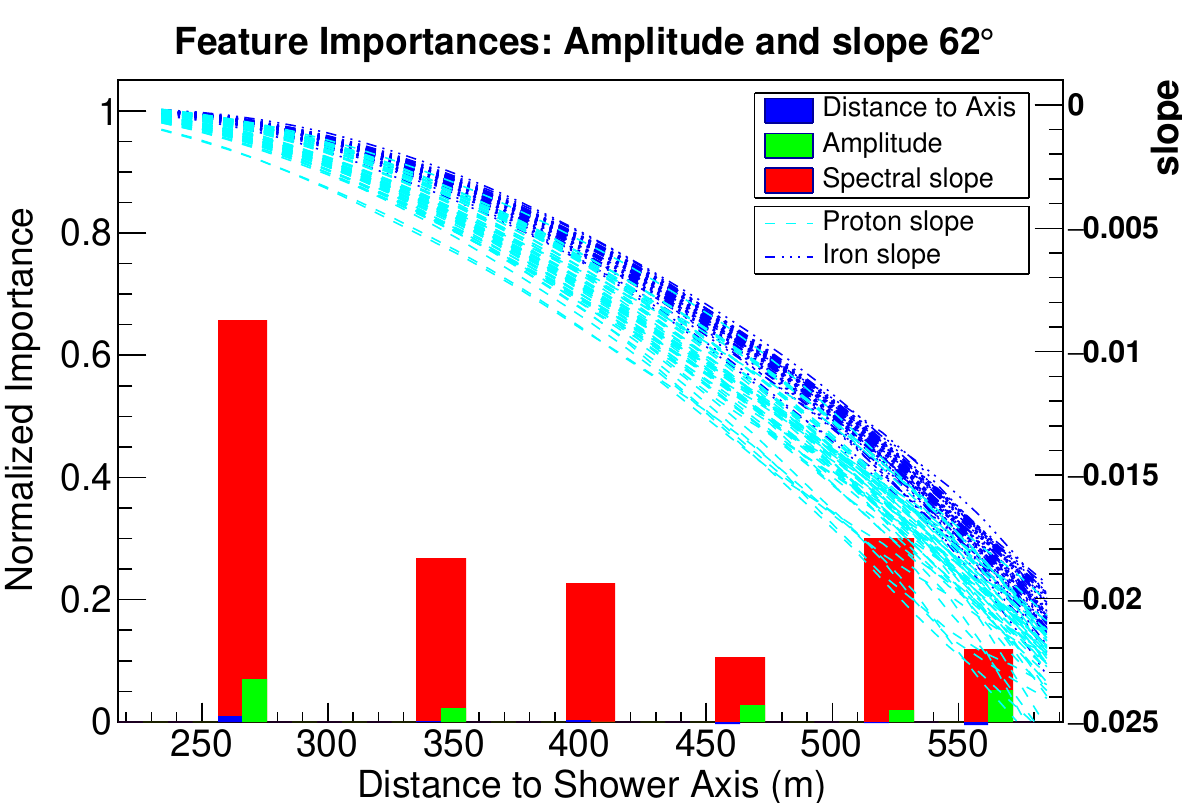}
   
  \end{center}
  
  
  \caption{Top: Feature importances at $\theta=82^\circ$, normalized to the most important feature, shown as a function of the average antenna distance to the shower axis. Overlaid are the spectral slopes obtained from the full ZHAireS simulations at the same zenith. Bottom left: Same as top, but for $\theta=62^\circ$ and with amplitudes overlaid instead. Bottom right: Same as left, but with spectral slopes overlaid.}
  
  \label{fig:Importances}
\end{figure}

To illustrate how the RF uses the available observables to perform the discrimination, Fig.~\ref{fig:Importances} shows the feature importances for two representative zenith angles. These importance patterns are shown only to provide qualitative insight into how the RF exploits the available observables. The importances are plotted as a function of the average distance to the shower axis associated with each feature. Overlaid on each panel we also show either the amplitude LDF or the spectral slope as a function of distance, as obtained from the full ZHAireS simulations. At $\theta = 82^\circ$ (top panel), the amplitude feature importances are negligible, suggesting that the RF only uses the slope for the discrimination. This is consistent with the behavior observed when using single observables, as the discrimination power of $SS$ appears to be much larger than that of $|E|$ at high zeniths. Furthermore, the slope importances tend to be highest in the region where the p–Fe slope separation is largest (see overlaid ZHAireS curves), as one would expect when discrimination is driven by the slope observable. At $\theta = 62^\circ$ (bottom panels), both observables seem to contribute to the RF discrimination, which is consistent with the similar accuracies obtained at this angle for the $|E|$-only and $SS$-only configurations. The most important feature overall is the amplitude at around 150 m (see bottom left panel), in the region where the p–Fe amplitude differences are most pronounced. In the mid-distance region, the slope features take over, coinciding with the region where the p–Fe slope separation is largest (see bottom right panel). At the largest distances, the amplitude becomes relevant again (see bottom left panel). One possible interpretation is that the RF could be using the size of the footprint for the discrimination, since proton showers tend to have smaller footprints. The high-importance regions shown in both panels for $|E|$ and $SS$ are consistent with what one would expect when using one or the other as the discriminator. Another example of a similar importance analysis, for a different $|E|$-only configuration, can also be found in Fig.~1 of~\cite{RLDF-submitted}. The importance patterns shown here serve only as illustrative examples of the physics behind how the RF uses the two observables. It is important to note that the accuracies presented in Fig.~\ref{fig:Results}, which are the main results of this work, remain unchanged regardless of how these importance patterns are interpreted.

The overall zenith dependence of our results reflects the complementarity of $|E|$ and $SS$ as discriminators: at low to mid zenith both the amplitude and the spectral slope carry significant mass information, while at high zenith the discrimination is dominated by the spectral slope. This complementarity underlies the consistently high accuracies obtained for the combined $|E|+SS$ configuration over the whole studied zenith range, which varied from 81\% to 96\%. These values should be regarded as conservative upper limits, since we applied a relatively large EM-energy uncertainty to the generated events and did not systematically optimize the RF hyperparameters. Furthermore, they are intended as a generic feasibility assessment rather than detector-specific performance estimates. This level of performance shows that the combined use of peak amplitudes and spectral slopes contains enough composition information to, in principle, enable robust event-by-event mass discrimination.

\section{Discussion}
\label{sec:Discussion}

This work evaluates the feasibility of event-by-event mass discrimination with a RF classifier trained on radio observables. As described in Section~\ref{sec:events}, the simulated events emulate those in a single EM-energy bin reconstructed using modern radio EM reconstruction methods and include a deliberately generous 10\% EM energy uncertainty. We now briefly discuss why the radio observables used here carry mass information, summarize the main limitations and scope of the analysis, comment on the use of fast simulation frameworks for large statistics ML training, and outline the most relevant directions for future development.

The mass sensitivity of the radio observables used in this work, $|E|$ and $SS$, comes from the fact that different primary compositions produce different \xmax distributions, which in turn affect the amplitude and spectral slope patterns at ground level. It is common to assume that the bulk of the radio emission comes from \xmax. An increase in \xmax reduces the geometric distance from it to the antennas, which tends to increase the observed amplitude. On the other hand, it also increases the density at the emission region, which in turn decreases the emission. The distance and density scalings are therefore competing effects and, together with coherence effects, create the composition-dependent amplitude behavior discussed in~\cite{RLDF-submitted}.

It is well known that the spectral slope depends on the shower development and therefore carries composition information~\cite{GrebeSS-ARENA2012,RossettoSS-ARENA2016,Huege2016PhysRep}. The dominant effect is geometrical in nature. When \xmax is farther from the antenna, a wider section of the shower development around \xmax contributes coherently to the measured signal. This larger region increases the spread in pulse arrival times, which results in a broader pulse and ultimately leads to a steeper spectral slope. Note that since the steepness of the spectrum also depends on the antenna position relative to the Cherenkov ring, this interpretation applies when comparing observers that view \xmax at approximately the same angle.

The results presented in Section~\ref{sec:Results} show a clear zenith-dependent behavior of the two radio observables used in this work. At low to mid zeniths, both the peak electric field amplitude $|E|$ and the spectral slope $SS$ provide significant discrimination power, as reflected by the comparable accuracies obtained when using either observable alone. At high zeniths, the discrimination obtained from $|E|$ rapidly degrades, while $SS$ accuracies increase and appear to dominate the RF decision. This picture is further corroborated by the feature importance analysis, which shows contributions from both observables at low to mid zeniths and an almost exclusive reliance on $SS$ at high zeniths.

We now summarize the main limitations and scope of the present analysis. The analysis already includes a deliberately generous EM shower energy uncertainty, which we expect to be one of the dominant experimental uncertainties, especially at low to mid zeniths, as it directly impacts the amplitude observable. However, no noise was included in our simulations. Given that this event-by-event discrimination method is geared towards high energies (high SNR), noise should have little effect on the peak amplitudes. At the same time, since galactic noise is much higher at low frequencies, it may slightly affect the spectral slope of the low signal antennas far from the Cherenkov ring. This could possibly diminish the slope discrimination power slightly, although the overall conclusions of this feasibility study are not expected to change.
We also did not include arrival direction uncertainties in the present analysis. In the regime of high SNR and for the dense array considered in this study, such uncertainties are expected to be well into the sub-degree regime (up to a few tenths of a degree, as demonstrated e.g. by LOFAR~\cite{LOFARAstropartDir,LOFARAngularError}). Nevertheless, direction errors can in principle modify the event geometry and thus may slightly smear the $|E|$ and $SS$ patterns, especially at extremely high zenith angles. Assessing this effect quantitatively is deferred to future work.
In addition, a single generic array geometry was used in this study, with an antenna spacing of 500 m. Alternative spacings were not explored in detail, although we performed simple tests with a larger spacing of 1 km, which led to a degradation of less than 10\% in discrimination accuracy at the lowest zenith angles, likely due to a poorer sampling of the radio footprint. At high zenith angles, the radio footprints are much larger and remain well sampled even with the sparser array.
Furthermore, this analysis was performed using a generic detector configuration. Although RDSim can incorporate detector details such as experiment-specific trigger conditions and antenna beam patterns, we intentionally did not use these options to keep the results more general. While beam patterns could in principle affect triggering, their impact on the observables used here is expected to be small, since these are typically reconstructed quantities corrected for detector response. The reported accuracies should therefore be interpreted as feasibility estimates within this intentionally simplified and generic setup, and not as performance forecasts for any specific experiment.
Finally, the discrimination accuracies obtained in this work reflect the specific set of full ZHAireS simulations used as input. Using a different set of simulated showers, even with the same parameters, could slightly modify the \xmax distributions and the associated amplitude and slope patterns, possibly resulting in small shifts in discrimination accuracy. This reflects a statistical uncertainty related to the finite simulation sample and cannot be quantified within the present study. This limitation is common to all simulation-based composition analyses. In the same vein, we only used a single hadronic model in our simulations (SIBYLL~\cite{SIBYLL}). Using a different hadronic model would also lead to different \xmax distributions and different discrimination accuracies. This hadronic model dependence was not explored in the present study, but we expect it to be larger than the statistical fluctuations described above.

The method presented here could, in principle, be applied to real data using the same analysis chain. However, since the RF is trained exclusively on simulations, any such application would be affected by potential differences between simulated and real events. Quantifying such differences and their impact on accuracies is beyond the scope of this study, as it would require experimental validation. This limitation is not specific to our method, but applies to all simulation-based composition analyses.

Only minimal empirical tuning was applied to the RF used in this work. The quoted accuracies therefore represent conservative estimates rather than maximally optimized values. Furthermore, the analysis was intentionally restricted to only two radio observables, $|E|$ and $SS$, in order to keep the method simple and physically interpretable. The inclusion of additional observables was not investigated in the present study, but could be explored in future work.

This feasibility study relies on RDSim to generate a large number of events from a limited set of ZHAireS simulations. This approach allows the use of large simulation statistics at a much lower computational cost compared to using full simulations alone. To the best of our knowledge, this work provides the first example of using a fast radio emission model to train an ML classifier for cosmic ray mass discrimination. The goal here is to demonstrate the feasibility of this approach, not to assess differences relative to training exclusively on full simulations. Although the peak amplitude and spectral slope observables we used in this work are typically derived from time traces, they are ultimately time-independent quantities once extracted. In RDSim, these observables are constructed at several positions along a reference line, where full ZHAireS time traces are available, and then incorporated into a time-independent emission model. The model then uses geometrical symmetries to estimate the same observables at any arbitrary antenna, without requiring the time trace at that particular position. With all these considerations in mind, we now discuss possible directions for expansion beyond the present feasibility scope.

A natural extension of this work is to go beyond the current proton–iron binary classification by including intermediate mass classes, such as a CNO group. This can be achieved within the same analysis framework we used here and would allow investigating how the discrimination performance changes as additional mass classes are introduced.

Another possible next step is to apply the method presented here to more detailed experimental setups. The incorporation of specific array geometries, trigger conditions, arrival direction uncertainties, antenna beam patterns, and noise models is already supported by RDSim. This would allow one to obtain, within a controlled and computationally efficient setup, more realistic experiment-dependent accuracy estimates that go beyond the upper limits presented in this feasibility work. We have previously discussed why we expect that including these additional detector effects will lead to only a modest degradation of the discrimination performance. However, in the event that a larger degradation is observed, the inclusion of additional radio observables that are less sensitive to these uncertainties could be explored to recover robustness. A further long-term extension is to evaluate the method using full ZHAireS simulations combined with a fully detailed detector response pipeline, instead of the fast RDSim framework. Initially, such studies could be limited to detector effects that are already available in RDSim. This would enable a direct comparison between the discrimination accuracies obtained with fast simulations and with full simulations, providing a quantitative assessment of how closely the results from these two approaches agree with each other. In a second step, additional detector uncertainties that cannot be modeled by RDSim, such as calibration uncertainties, electronics response, and realistic environmental noise, could be included to compute more realistic discrimination accuracies. However, this would require access to detailed detector information and validated reconstruction pipelines, and would therefore only be possible at a collaboration level.

Finally, a much more ambitious long-term direction is to study the agreement between simulated radio observables and those reconstructed from real events. Although event-by-event composition cannot be verified experimentally, hybrid events with independently reconstructed geometry and \xmax would allow direct comparisons between reconstructed amplitudes and spectral slopes from real events with their simulated counterparts. Such comparisons would help quantify simulation–reality differences at the level of radio observables and could also guide the choice of simulation settings that best reproduce the data, including the choice of hadronic model. Such a study would require large hybrid data sets and matched reconstruction frameworks, and would therefore constitute a large collaboration-level effort.

\section{Conclusions}
\label{sec:Conclusions}

This work investigates the feasibility of event-by-event primary mass discrimination using radio observables. The analysis is performed using peak electric-field amplitudes and spectral slopes of the radio signal and does not require an explicit reconstruction of \xmax. Although the shower maximum is not reconstructed, the discrimination power still arises from the sensitivity of these observables to the longitudinal development of the extensive air shower. To assess this feasibility, a supervised ML classifier is used.

The discrimination performance still shows a clear dependence on zenith angle, even when the peak amplitudes and the spectral slope are used simultaneously. At low to mid zeniths, both $|E|$ and $SS$ contribute significantly to the discrimination, while at high zenith angles the discrimination is dominated by $SS$.
The complementary nature of these two observables results in a stable discrimination power across the full zenith range studied, where we obtained accuracies varying from 81\% to 96\%, even after normalizing each shower by its own electromagnetic energy. Since the analysis includes deliberately conservative choices, such as a large 10\% uncertainty on the reconstructed EM energy, these accuracies should be interpreted as conservative upper limits. Our results show that event-by-event mass discrimination using these radio observables is, in principle, feasible.

This work is intentionally limited to a generic feasibility assessment and does not provide performance estimates for any specific detector, as this would require the inclusion of additional detector-specific effects not considered here. Examples of such effects include noise, arrival-direction uncertainties, and antenna beam patterns. Incorporating these effects can be achieved within the same simulation and analysis framework used in this work, and would effectively transform the present feasibility study into a detector-specific performance evaluation. As pointed out in the Discussion, since what we believe to be the dominant uncertainty affecting radio amplitudes is already included, we do not expect the addition of these detector-specific effects to qualitatively change our feasibility conclusions.

In this work, we used fast radio emission simulations coupled with a simplified detector response model to generate the event samples required for training the RF classifier. To the best of our knowledge, this work provides a first example of using fast radio simulations to train ML models for cosmic ray mass composition studies, illustrating that large training samples can be obtained and used to train ML models without relying on extensive sets of full Monte Carlo simulations. No attempt was made to train machine learning models using full Monte Carlo simulations alone, and therefore no comparison between the two training approaches is presented.

The discrimination accuracies reported in this work were obtained using an RF classifier to perform a binary classification into light and heavy primary mass classes. As input features, it exclusively relied on the peak electric field, the spectral slope, and the distance to the shower axis for each triggered antenna. The conservative upper limits obtained for the discrimination accuracies show that, in principle, event-by-event primary mass discrimination using radio observables is feasible.

\section{Acknowledgements}
We would like to thank Jaime Alvarez-Muñiz for the many suggestions and proof reading the manuscript. We acknowledge the Polish National Agency for Academic Exchange within Polish Returns Program no. PPN/PPO/2020/1/00024/U/00001,174; National Science Centre Poland for NCN OPUS grant no. 2022/45/B/ST2/0288. Disclosure: During the final editing of the manuscript, we used the OpenAI language model ChatGPT (GPT-5) as a tool to assist with checking grammar, typos, clarity and bibliographic formatting. All text was written and revised by the authors, who take full responsibility for the content of the article.

\section{Appendix}
\label{sec:Appendix}

This Appendix provides illustrative examples and technical details related to the RDSim framework and to the extraction of the spectral slope (SS) radio observable used in the machine-learning analysis. The material shown here complements the main text by illustrating the effects of the azimuthal rotation of the emission model, presenting additional comparisons with full shower and detector simulations, detailing the extraction of the spectral slope from frequency spectra, and showing an example RDSim event.

\subsection{Comparison between dedicated and rotated emission models}
\label{sec:A-RotationExample}

RDSim allows the radio emission model constructed from a ZHAireS simulation to be rotated in azimuth. This enables the generation of events with different arrival directions from a single ZHAireS simulation, while accounting for the corresponding change in the angle between the shower axis and the geomagnetic field. The general rotation procedure is described in Section~\ref{sec:toymodel}. In this approach, the emission model is rotated as a whole, including the reference line and the associated ellipses defining the footprint, and no interpolation in azimuth is performed. The Askaryan and geomagnetic components are treated separately, with only the geomagnetic contribution rescaled to account for the change in geomagnetic angle.

\begin{figure}[!htb]
  \begin{center}
    
     \includegraphics[width=\linewidth]{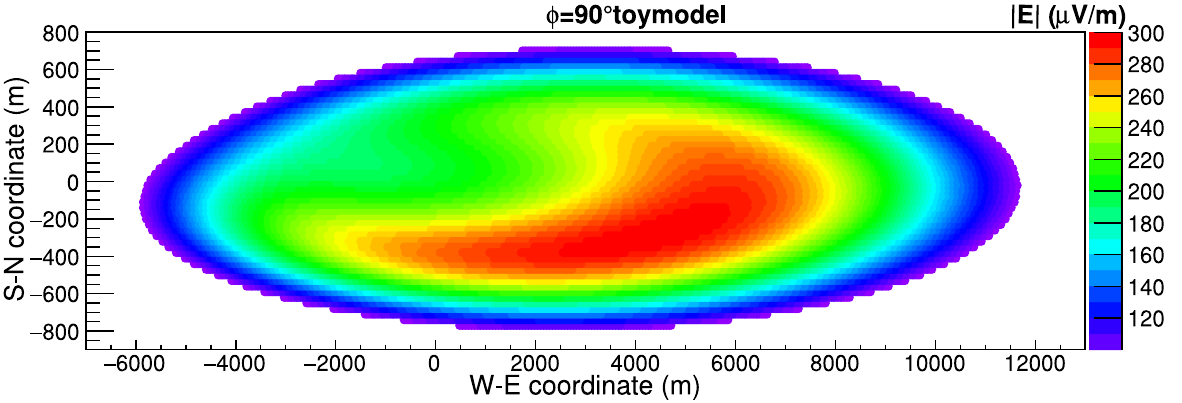}
    \includegraphics[width=\linewidth]{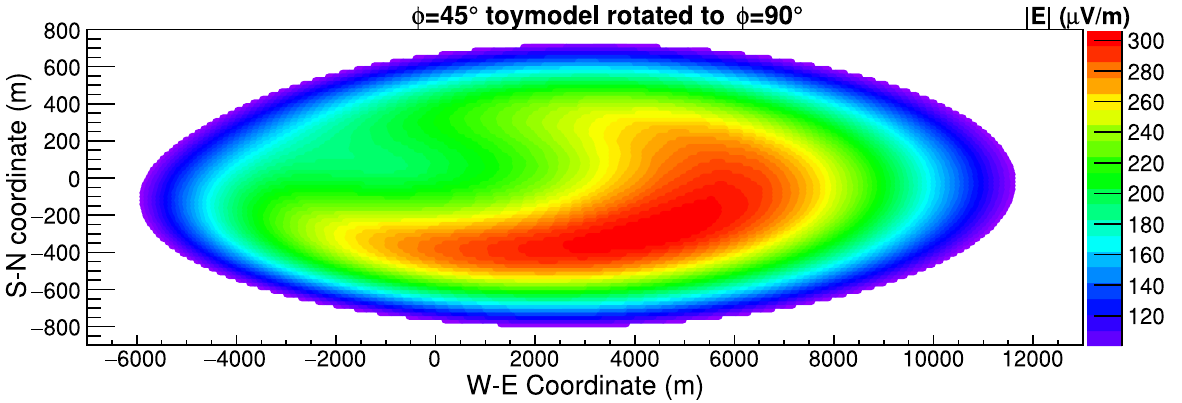}
  \end{center}
  \vspace{-0.5cm}
  \caption{Comparison between a dedicated emission model instance coming from the West (top) and a rotated emission model instance, originally constructed using a simulation of a shower coming from the NW (bottom). The maximum difference observed in this example comparison is 2\%.}
  \label{fig:toyrotation}
\end{figure}

An example comparison illustrating the effect of this procedure is shown in Fig.~\ref{fig:toyrotation}. The figure compares a dedicated emission model constructed from a simulation already at the target azimuth of $\phi = 90^\circ$ (top) with a rotated model obtained from a simulation initially at $\phi = 45^\circ$ (bottom). The two footprints exhibit very similar shapes and structures, with only small differences in absolute amplitudes, as indicated by the color scale. The maximum difference observed in this example comparison is 2\%. This example illustrates the typical level of differences introduced by the azimuthal rotation procedure. In the analysis presented in this work, the emission model is rotated over the full azimuth range. While only a representative example is shown here, similar differences are observed for other rotation angles, allowing the model to be rotated in practice without introducing large additional differences.

\subsection{Extraction of the spectral slope using frequency spectrum fits}
\label{sec:A-SSFit}

The spectral slope is used in this work as an input feature for the RF analysis. It is obtained from the frequency spectrum of the radio signal, which is calculated from the full electric field time traces available in the input ZHAireS simulations for antennas along the reference line. For each antenna, this quantity is extracted by fitting the logarithm of the frequency spectrum, $\log_{10}\!|E(f)|$, in the relevant frequency band.

For this fit, three simple polynomial models are implemented in the analysis and are defined in Eqs.~(\ref{eq:SSlinNoF0})–(\ref{eq:SSquadF0}): a linear fit without a reference frequency (Eq.~(\ref{eq:SSlinNoF0})), a linear fit around a fixed reference frequency $f_0$ (Eq.~(\ref{eq:SSlinF0})), and a second-degree polynomial fit around the same $f_0$ (Eq.~(\ref{eq:SSquadF0})). In all cases, the fit parameter labeled $\mathrm{slope}$  is used as the spectral slope feature SS in the RF analysis. The additional quadratic parameter, denoted as $m_{f2}$ in the second-degree fit, is included only to assess possible curvature of the spectrum and is not used further in the analysis. 
\begin{equation}
\log_{10}\!|E(f)| = A + \mathrm{slope}\, f
\label{eq:SSlinNoF0}
\end{equation}

\begin{equation}
\log_{10}\!|E(f)| = A + \mathrm{slope}\,(f - f_0)
\label{eq:SSlinF0}
\end{equation}

\begin{equation}
\log_{10}\!|E(f)|
=
A + \mathrm{slope}\,(f - f_0) + m_{f2}\,(f - f_0)^2
\label{eq:SSquadF0}
\end{equation}

\begin{figure}[!htb]
  \begin{center}
    \includegraphics[width=0.5\linewidth]{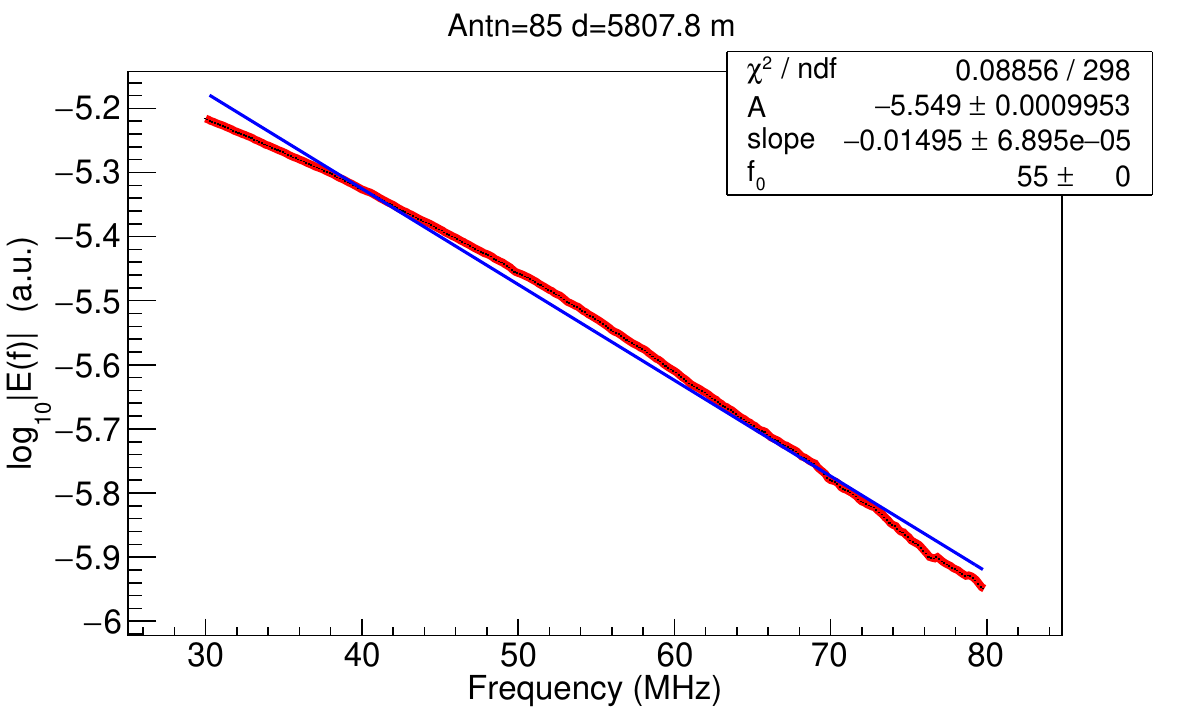}\includegraphics[width=0.5\linewidth]{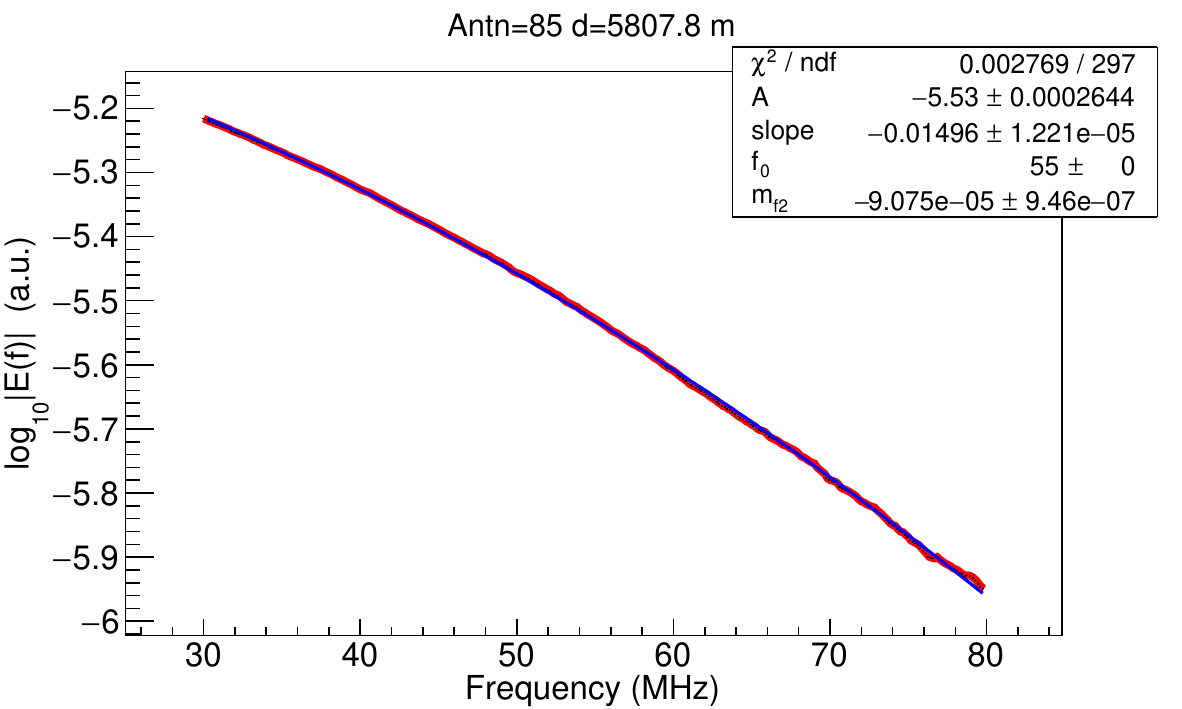}
  \end{center}
  \caption{Example fits used to extract the spectral slope (SS) from the radio-frequency spectrum at a single antenna. Shown are fits to $\log_{10}\!|E(f)|$ as a function of frequency for the same spectrum, using a linear model (left) and a quadratic model (right). In both cases, the spectral slope SS is defined as the fit parameter $\mathrm{slope}$. The additional quadratic parameter ($m_{f2}$) in the right panel is fitted only to assess possible curvature and is not used in the analysis. The extracted values of $\mathrm{slope}$ are essentially identical for both fit options.}
  \label{fig:FitExamples}
\end{figure}

We find that the value of the fit parameter $\mathrm{slope}$, which defines the spectral slope SS used in the analysis, is essentially insensitive to the specific fit option employed. Linear and quadratic fits, as well as fits with or without a reference frequency, yield nearly identical values. This robustness is illustrated by the example fits shown in Fig.~\ref{fig:FitExamples}. For the results presented in this work, only the quadratic fit was used in the analysis. Given the observed robustness of the extracted spectral slope, this choice does not affect the results.

\subsection{Comparison of RDSim with full shower and detector simulations}
\label{sec:A-Offline}

In this section we present illustrative example comparisons between events generated with RDSim, including both the radio emission and the simplified detector response, and corresponding events simulated using full air shower simulations combined with a complete detector response chain (PAO Offline). The purpose of these examples is not to provide a systematic validation of the framework, but rather to illustrate the level of agreement typically achieved by the fast RDSim approach at the detector level. The example comparisons shown here were originally published in~\cite{RDSim-ICRC2023}, and are included here for completeness, as referenced in the main text.

\begin{figure}[!htb]
  \begin{center}
    \hspace*{-1em}\includegraphics[width=0.35\linewidth]{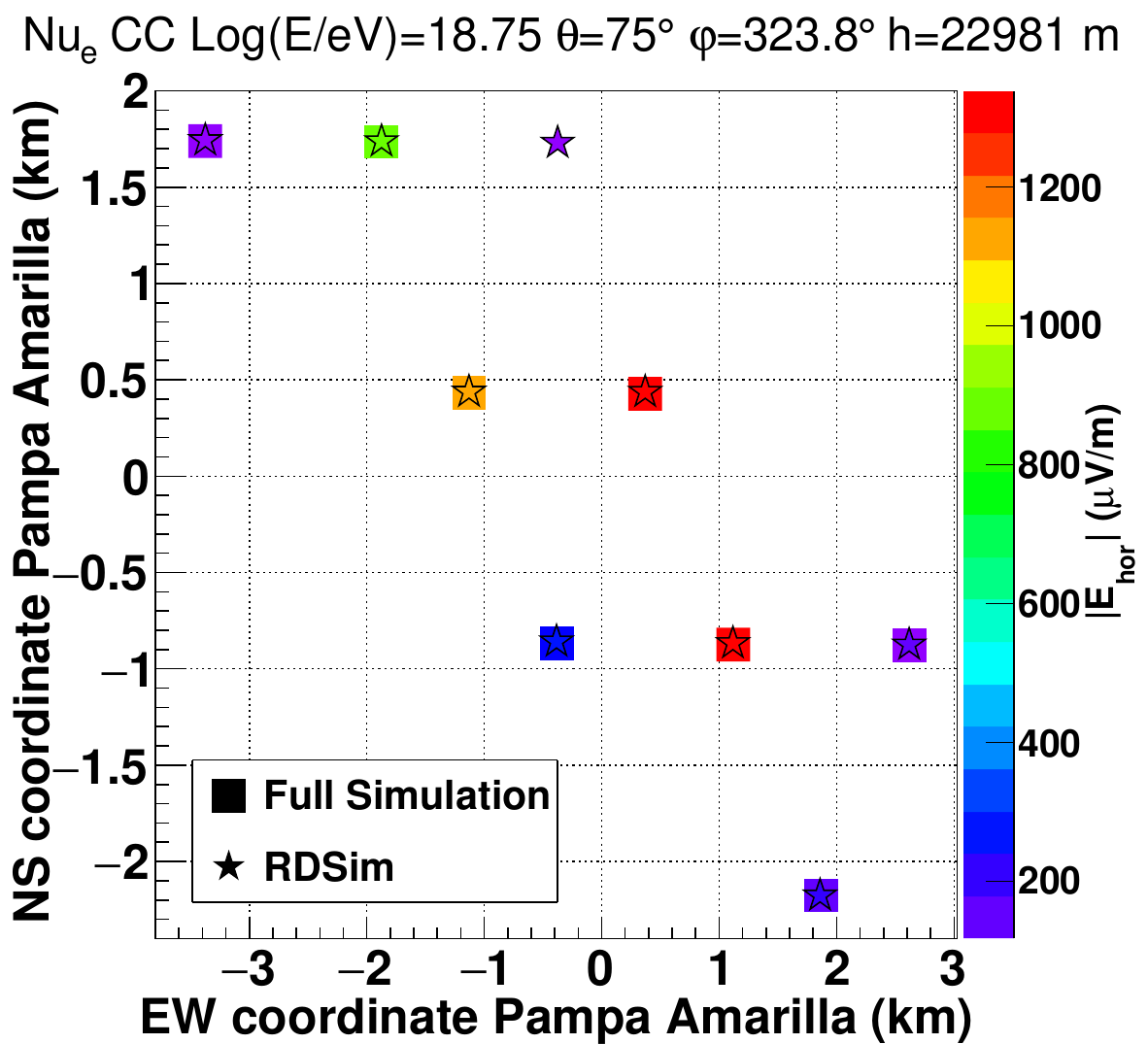}\includegraphics[width=0.35\linewidth]{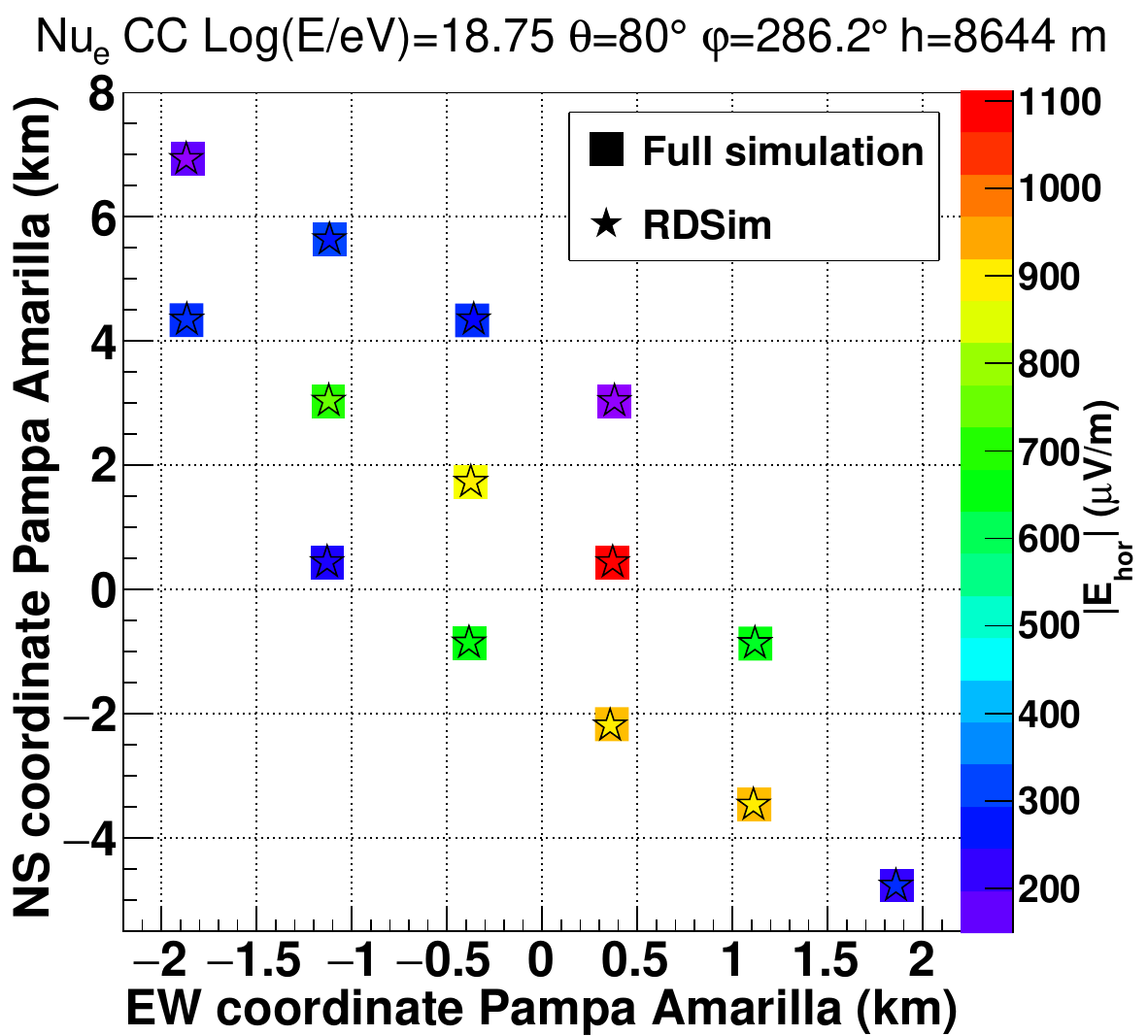}\includegraphics[width=0.35\linewidth]{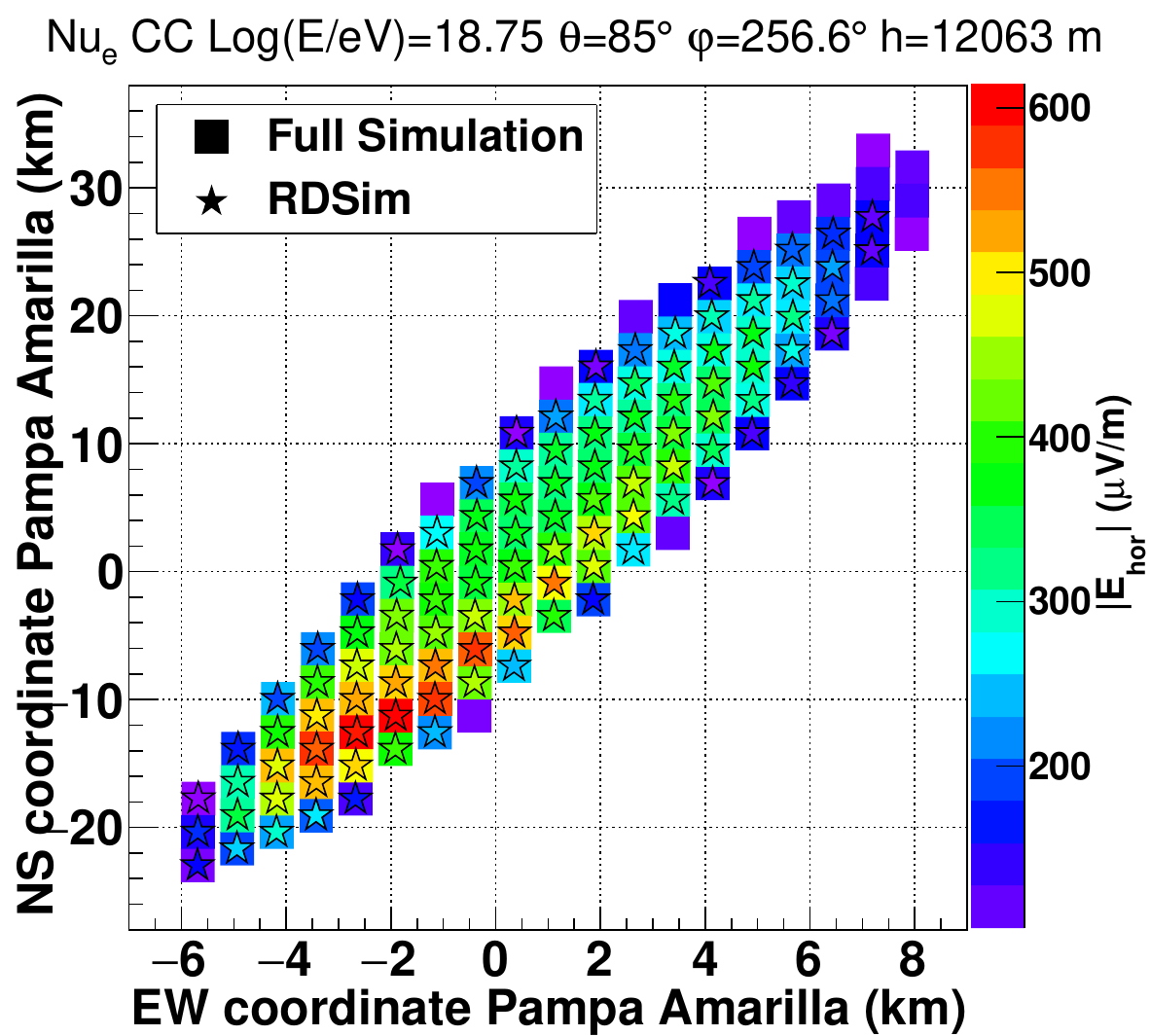}
  \end{center}
  \caption{Comparison of events induced by $\nu_e$ CC interactions in the atmosphere simulated by RDSim (stars), with the same events simulated using full detailed simulations of both the shower and the Auger Radio Detector (squares). The zenith angles of the events are 75$^\circ$ (left), 80$^\circ$ (middle) and 85$^\circ$ (right). The color scales inside the star and inside the square represent the electric field at each antenna, as obtained from RDSim and the full simulation, respectively.}
  \label{fig:offlinecomp}
\end{figure}

For each example shown in Fig.~\ref{fig:offlinecomp}, the same underlying shower geometry and shower development are used to generate two event representations. In the first case, the radio emission is modeled using RDSim and passed through its simplified detector response. In the second case, the full air shower radio emission is simulated with ZHAireS, with the electric field time trace calculated at each antenna position, and subsequently processed through the complete PAO Offline detector simulation chain. The resulting events are then compared at the detector level in terms of the reconstructed peak electric field amplitudes at each triggered antenna position.

Figure~\ref{fig:offlinecomp} compares the reconstructed peak electric field footprints obtained with RDSim (stars) and with the full simulation and detector response chain (squares) for three different highly inclined events. In all cases, the spatial distribution of triggered antennas and the relative amplitude patterns across the array show very good overall agreement between the two approaches. Differences are visible both in the absolute amplitudes and in the exact set of antennas that trigger, especially at the highest zenith angle (right panel), where some antennas trigger in the full simulation but not in RDSim. Nevertheless, despite the simplified detector response used by RDSim, the main detector level characteristics of the radio footprints are consistently reproduced.

These example comparisons are intended to be illustrative and are not meant to constitute a comprehensive or systematic validation of the RDSim framework. Nevertheless, the level of agreement observed at the detector level is sufficient for the purposes of the present feasibility work. In particular, the simplified detector response implemented in RDSim captures the dominant features relevant for the large-scale generation of training samples used in the ML analysis.

\subsection{Example RDSim event}
\label{sec:A-EventExample}

\begin{figure}[!htb]
  \begin{center}
    \includegraphics[width=0.5\linewidth]{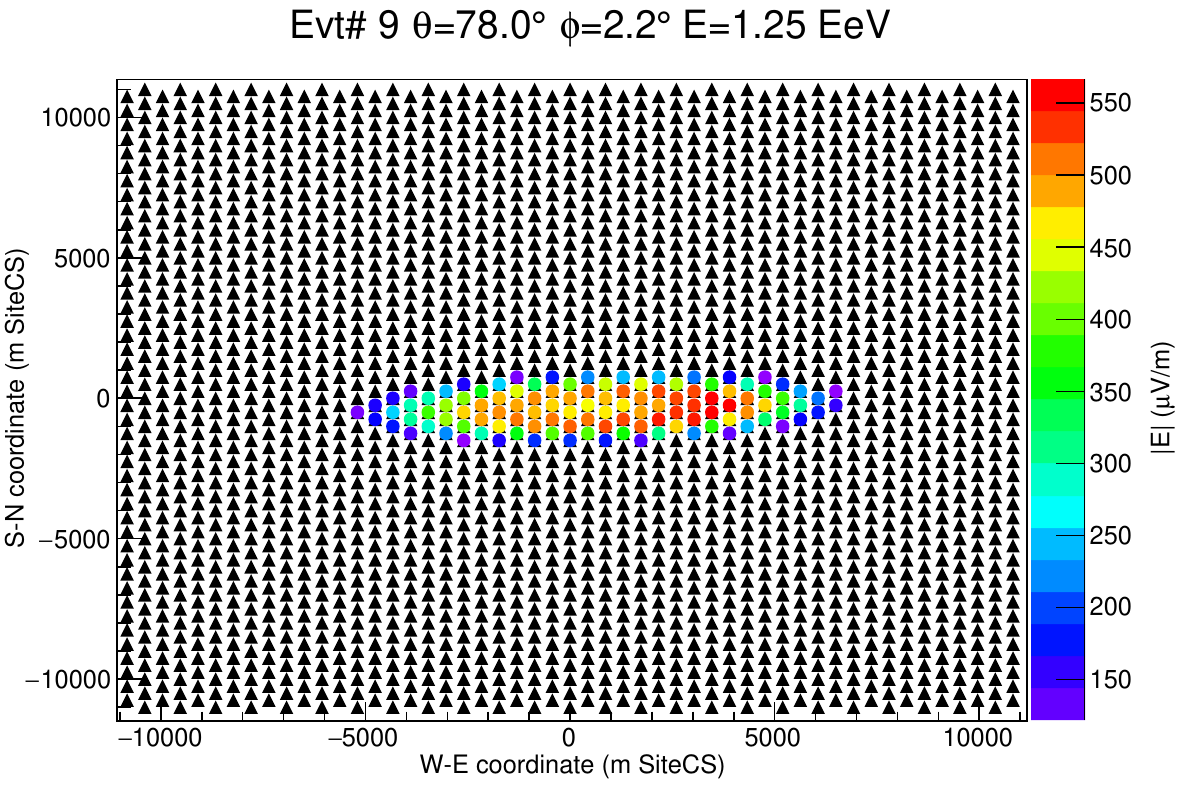}\includegraphics[width=0.5\linewidth]{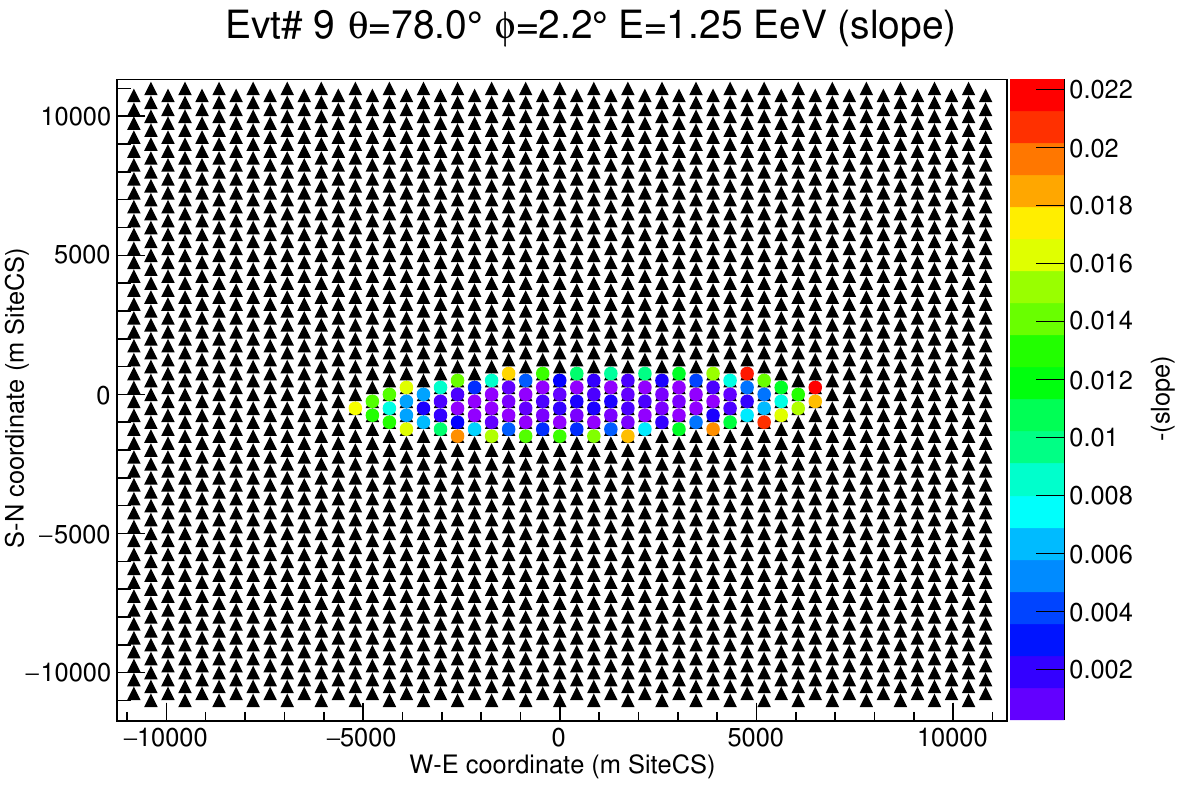}
  \end{center}
  \caption{Example RDSim event at a zenith angle of $78^\circ$ showing the peak electric field amplitude at ground level (left) and the corresponding spectral slope distribution (right). For visualization purposes, the sign of the fitted slope is inverted, such that larger values correspond to steeper frequency spectra.}
  \label{fig:EventExample}
\end{figure}

An illustrative example of a single RDSim generated event is shown in Fig. \ref{fig:EventExample} for a zenith angle of $78^\circ$. The left panel shows the peak electric field amplitude at ground level, while the right panel shows the corresponding spectral slope (SS) distribution obtained for the same event. Both quantities are evaluated at each triggered antenna and correspond to the radio observables used as input features in the RF analysis. This example is included solely to provide a concrete illustration of a typical radio footprint and SS pattern, and is not intended to demonstrate any discrimination power or performance. The plots shown here were automatically generated by the RDSim framework using its optional event plotting functionality.

\end{document}